\documentclass[aps,prl,preprint,superscriptaddress]{revtex4-2}

\usepackage{chemformula} 
\usepackage[T1]{fontenc} 

\usepackage{eufrak}
\usepackage{amsmath}
\usepackage{mathrsfs}
\usepackage{mathtools}
\usepackage{multirow}
\usepackage{amsfonts}
\usepackage{bm}
\usepackage{bbold}
\usepackage{multirow}

\newcommand*{\ket}[1]{\ensuremath{\left| #1 \right>}}
\newcommand*{\ketbig}[1]{\ensuremath{\big| #1 \big>}}

\newcommand*{\bracket}[2]{\ensuremath{\left< #1 \big| #2 \right>}}
\newcommand*{\bracketbig}[2]{\ensuremath{\big< #1 \big| #2 \big>}}

\newcommand*{\ketbra}[2]{\ensuremath{\left| #1 \big>\big< #2 \right|}}

\newcommand*{\average}[3]{\ensuremath{\left< #1 \big| #2 \big| #3 \right>}}

\newcommand*{\projective}{\ensuremath{\mathbb{P}}}
\newcommand*{\real}{\ensuremath{\mathbb{R}}}

\newcommand*{\orbSp}{\ensuremath{\mathcal{W}}}
\newcommand*{\grass}[2]{\ensuremath{\text{Gr}(#1,#2)}}
\newcommand*{\stiefel}[2]{\ensuremath{\text{ST}(#1,#2)}}
\newcommand*{\extAlg}{\ensuremath{\bigwedge \orbSp}}
\newcommand*{\extProd}{\ensuremath{\bigwedge\nolimits^n \orbSp}}
\newcommand*{\extProdTwo}{\ensuremath{\bigwedge\nolimits^2 \orbSp}}

\newcommand*{\w}{\wedge}

\newcommand*{\HF}{\ensuremath{\Phi_{\text{HF}}}}
\newcommand*{\minD}{\ensuremath{\Phi_{\text{minD}}}}
\newcommand*{\extWF}{\ensuremath{\Psi_{\text{ext}}}}
\newcommand*{\CI}{\ensuremath{\Psi_{\text{CISD}}}}
\newcommand*{\CC}{\ensuremath{\Psi_{\text{CCSD}}}}

\newcommand*{\Kmat}{\ensuremath{\mathbf{K}}}
\newcommand*{\Hess}{\ensuremath{\mathfrak{H}}}
\newcommand*{\Jac}{\ensuremath{\mathfrak{J}}}
\newcommand*{\Habsil}{\ensuremath{\mathcal{H}}}
\newcommand*{\Jabsil}{\ensuremath{\mathcal{J}}}

\newcommand*{\matG}{\ensuremath{{\mathbf{G}}}}
\newcommand*{\matM}{\ensuremath{{\bm{\mathcal{M}}}}}
\newcommand*{\bigG}{\ensuremath{{\mathscr{G}}}}
\newcommand*{\bigH}{\ensuremath{{\mathscr{H}}}}

\newcommand*{\Dss}{\ensuremath{\mathcal{C}}}
\newcommand*{\Dmixaa}{\ensuremath{\mathcal{A}}}
\newcommand*{\Dmixab}{\ensuremath{\mathcal{B}}}
\newcommand*{\Dmix}{\ensuremath{\mathcal{D}}}

\newcommand*{\irp}{\ensuremath{\Gamma}}
\newcommand*{\irpP}{{\ensuremath{\Gamma'}}}
\newcommand*{\irpPP}{{\ensuremath{\Gamma''}}}
\newcommand*{\irpPPP}{{\ensuremath{\Gamma'''}}}
\newcommand*{\irpB}{{\ensuremath{\overline{\Gamma}}}}
\newcommand*{\irpBB}{{\ensuremath{\overline{\overline{\Gamma}}}}}

\newcommand*{\sumijabrestr}
{\ensuremath{\quad\sum_{\mathclap{\left.\substack{(i>j)\\(a>b)}\right\}\in\irp}}\quad}}
\newcommand*{\sumijabfull}
{\ensuremath{\quad\sum_{\mathclap{\left.\substack{(i,a)\\(j,b)}\right\}\in\irp}}\quad}}
\newcommand*{\sumijabmix}
{\ensuremath{\quad\sum_{\mathclap{\substack{(i,a)\in\irp\\(j,b)\in\irpP}}}\quad}}

\newcommand*{\sumB}{\ensuremath{\sum_{\irpB \ne \irp}}}
\newcommand*{\sumBB}
{\ensuremath{\sum_{\mathclap{\substack{\irpBB > \irpB\\\irpB \ne \irp\\\irpBB \ne \irp}}}}}
\newcommand*{\sumBp}{\ensuremath{\sum_{\irpB \ne \irp,\irpP}}}
\newcommand*{\sumBBp}
{\ensuremath{\sum_{\mathclap{\substack{\irpBB > \irpB\\\irpB \ne \irp,\irpP\\\irpBB \ne \irp,\irpP}}}}}

\begin{document}


\title{Calculating the distance from an electronic wave function to
  the manifold of Slater determinants through the geometry of Grassmannians}

\author{Yuri Alexandre Aoto}
\email{yuri.aoto@ufabc.edu.br}
\affiliation
{Center for Mathematics, Computing and Cognition,
  Federal University of ABC (UFABC),
  Santo Andr{\'e}, 09210-580 S{\~a}o Paulo, Brazil}
\author{M{\'a}rcio Fabiano da Silva}
\affiliation
{Center for Mathematics, Computing and Cognition,
  Federal University of ABC (UFABC),
  Santo Andr{\'e}, 09210-580 S{\~a}o Paulo, Brazil}

\date{\today}

\begin{center}
  This material, which the American Physical Society has the copyright,\\
  has been originally published at Phys. Rev. A  102, 052803\\
  \url{https://journals.aps.org/pra/abstract/10.1103/PhysRevA.102.052803}\\
   
\end{center}

\begin{abstract}
  The set of all electronic states that can be expressed as a single Slater determinant
  forms a submanifold, isomorphic to the Grassmannian,
  of the projective Hilbert space of wave functions.
  We explored this fact
  by using tools of Riemannian geometry of Grassmannians as described by Absil et. al
  [Acta App. Math. \textbf{80}, 199 (2004)],
  to propose an algorithm that converges to a Slater determinant that is critical point
  of the overlap function with a correlated wave function.
  This algorithm can be applied to quantify the entanglement or correlation of a wave function.
  We show that this algorithm is equivalent to the Newton method using the standard parametrization
  of Slater determinants by orbital rotations,
  but it can be more efficiently implemented because the orbital basis used to express the
  correlated wave function is kept fixed throughout the iterations.
  We present the equations of this method for a general configuration interaction wave function
  and for a wave function with up to double excitations over a reference determinant.
  Applications of this algorithm to selected electronic systems are also presented and discussed.
\end{abstract}

\maketitle

\section{\label{sec:intro}Introduction}


Electron correlation is at the heart of electronic structure theory,
and its intimate relation to quantum entanglement, as viewed by quantum information theory,
attracts the attention of researchers from both fields
\cite{gersdorfIJQC97_61_935,benavides-riverosPRA17_95_032507,dingJCTC20_16_4159}.
From the point of view of atomic and molecular physics,
the effect of electron correlation on the electronic energy is the most important feature to be considered,
although its consequences on properties are also relevant in several applications.
From the side of quantum information theory, one is often interested in quantifying the entanglement of a wave function intrinsically \cite{horodeckiRMP09_81_865,myersQIP10_9_239,dingJCTC20_16_4159},
irrespective of any observable, in particular the energy.
Many ways to measure entanglement have been proposed \cite{horodeckiRMP09_81_865},
for instance, by the distance between the quantum state and the set of states with no entanglement (uncorrelated states).
This definition has a geometric nature, and calls for the geometry of the sets of quantum states.


In wave function methods of electronic structure,
the correlation due to the fermionic character of the electrons is always taken into account by using anti-symmetrized wave functions.
Slater determinants, which represent mean-field states, are the simplest of such wave functions.
The true ground state wave function, on the other hand, presents extra electron correlation apart of that associated with the Pauli principle,
and if one is concerned with measuring this extra correlation/entanglement in the wave function,
its distance to the set of Slater determinants is an expected approach
\cite{benavides-riverosPRA17_95_032507}.
Using the set of configuration state functions, that are spin eigenfunctions,
is a similar possibility, although more involved.


\newcommand*{\PsiA}{\Psi}
\newcommand*{\PsiB}{\Psi'}
There are multiple ways to define a metric in the space of electronic states.
It cannot be a metric on the Hilbert space of wave functions,
but on its projective space instead,
as it must reflect the fact that the normalization and phase of the wave functions do not alter their associated physical states.
Some examples of metrics used in quantum mechanics are:
\begin{equation}\label{eq:metric_FubiniStudy}
  D_\text{FS}(\PsiA,\PsiB) = \text{arccos}|\bracketbig{\PsiA}{\PsiB}|
\end{equation}
\begin{equation}\label{eq:metric_Damico}
  D_\text{ACFC}(\PsiA,\PsiB) = \sqrt{1 - |\bracketbig{\PsiA}{\PsiB} |}
\end{equation}
\begin{equation}\label{eq:metric_simple}
  D_\text{BRLCM}(\PsiA,\PsiB) = 1 - |\bracketbig{\PsiA}{\PsiB}|^2\,,
\end{equation}
where the wave functions $\PsiA$ and $\PsiB$ are assumed to be normalized to unit.
Equation~\eqref{eq:metric_FubiniStudy} is the Fubini-Study metric,
introduced in quantum mechanics by Bures \cite{buresTAMS69_135_199,hubnerPLA92_163_239},
and it can be interpreted as the angle between both state vectors.
The second equation has been studied by D'Amico et. al 
\cite{damicoPRL11_106_050401}, together with a related metric in the space of densities
(their original metric uses a different normalization condition).
The last one was recently used by Benavides-Riveros et. al
\cite{benavides-riverosPRA17_95_032507},
who provided an upper bound to $D_\text{BRLCM}(\HF, \Psi_0)$
based on the correlation energy (for systems with non-degenerate ground state):
\begin{eqnarray}\label{eq:BRLCM_upper_bound}\nonumber
  1 - |\bracketbig{\HF}{\Psi_0}|^2
  &=& D_\text{BRLCM}(\HF, \Psi_0)\\
  &\le& \frac{|E_\text{corr}|}{E_\text{gap}}
  = \frac{E_\text{HF} - E_0}{E_1 - E_0}\,,
\end{eqnarray}
where $\HF$ and $\Psi_0$ are the Hartree-Fock and the exact ground state wave functions, respectively, and $E_0$, $E_1$, and $E_\text{HF}$ are the ground state-, first excited state-, and Hartree-Fock energies.
Furthermore, quantification of static/dynamic correlation incorporated by a wave function based on this metric has also been proposed \cite{benavides-riverosPCCP17_19_12655}.


For all the above metrics,
the distance between two states described by wave functions $\PsiA$ and $\PsiB$ is related to the absolute value of their overlap,
$|\bracketbig{\PsiA}{\PsiB}|$, by Equations \eqref{eq:metric_FubiniStudy} to \eqref{eq:metric_simple}.
Therefore, calculating the distance from a given correlated electronic wave function to the set of Slater determinants is equivalent to finding the Slater determinant that maximizes the overlap with such wave function
(constrained to normalized wave functions).
This is not a trivial task \cite{dingJCTC20_16_4159},
and analytical expressions are known only for the small case of two-particle systems \cite{zhangPRA14_89_012504}
and for some specific cases \cite{zhangPRA16_94_032513}.
In \cite{benavides-riverosPRA17_95_032507}, Benavides-Riveros et. al opted to measure the correlation of the ground state wave function by its distance to the Hartree-Fock wave function (in Eq.~\eqref{eq:BRLCM_upper_bound}), that is not, in general, the minimizer of the distance among all possible Slater determinants.
Since no analytical procedure is known for the general case,
finding the Slater determinant that minimizes the distance to a correlated wave function requires a numerical optimization on the space of Slater determinants.
This optimization problem has been studied by Zhang and Kollar \cite{zhangPRA14_89_012504} and by Zhang and Mauser \cite{zhangPRA16_94_032513},
by both analytical and numerical procedures.
They have also provided an algorithm that converges monotonically,
but slowly, to the Slater determinant with the largest overlap with a correlated wave function.
Their algorithm is discussed below in Sec.~\ref{sec:comp_Zhang_Kollar}.


To fully appreciate the phenomenon of correlation in electronic structure,
one has to consider how the manifold of Slater determinants is embedded in the set of all electronic wave functions.
This manifold is the \emph{Grassmannian} \cite{borisenkoRMS91_46_45,baralicTTOM11_XIV_147}.
Although Slater determinants are of paramount importance in electronic structure theory
and Grassmannians are of high importance to geometry \cite{griffiths-78_0_0,hodge-94_1_0},
their connection is rarely observed when looking at Slater determinants.
The first work to establish this connection dates back to 1980, by Rowe, Ryman, and Rosensteel
\cite{rowePRA80_22_2363}.
Afterwards, Cassam-Chena{\"i} (in 1994) \cite{cassam-chenaiJMC94_15_303},
Panin (in 2007) \cite{paninA07_0_1},
and Chiumiento and Melgaard (in 2012) \cite{chiumientoJGP12_62_1866} have also studied the geometry of the Grassmannian in such context.
Very recently, Polack and coworkers have used the Grassmannian to formulate a procedure to obtain an initial guess for self-consistent field calculations \cite{polackMP20_0_0}.
However, systematic applications of the properties of this ``fundamental family of compact complex manifolds'' \cite{griffiths-78_0_0} to the electronic structure theory is still missing.


The objective of this article is to explore how the geometry of the Grassmannian can be used to perform the optimization
of the Slater determinant with the largest overlap with an arbitrary wave function,
and thus ultimately measuring its correlation.
We will show that if we explicitly consider the geometry of the Grassmannian
a more efficient algorithm can be obtained.
Moreover,
this algorithm is a Newton method that uses a set of \emph{non independent parameters} to describe the Slater determinants
(namely the coefficients matrices on a fixed orbital basis),
contrary to the usual assumption that a set of independent parameters
is necessary to carry out orbital optimizations with the Newton method.

In Sec.~\ref{sec:grassmannian} we review the relation between the Grassmannian and the field of electronic structure,
along with the Pl{\"u}cker embedding and the description of many-electron wave functions by the exterior algebra \cite{ruiz-tolosa-05_0_0,bowen-08_0_0}.
Up to our knowledge, researchers in molecular physics and theoretical chemistry usually have no familiarity with these concepts.
For the detailed treatment of electronic wave functions within the framework of exterior algebra the reader is referred to the works of Mundin \cite{mundimJPF89_50_11,mundimRBEF97_19_209},
and Cassam-Chena{\"i} \cite{cassam-chenaiJMC94_15_303}.
See also \cite{vourdasJPAMT18_51_445301} for applications of this approach in the context of quantum computation.
Sec.~\ref{sec:grassmannian} is ended with the mathematical formulation of the problem we will be concerned with.
Sec.~\ref{sec:algorithms} describes the algorithms we propose and compares them
from the theoretical point of view.
In Sec.~\ref{sec:examples} some numerical examples are discussed.
After the concluding remarks, we present in the Appendix the complete expressions for these algorithms over symmetry adapted spatial orbitals.

\section{\label{sec:grassmannian}Grassmannian and the Pl{\"u}cker embedding}


For a molecular system of $n$ electrons,
a finite-dimensional approximation for the space of wave functions can be obtained by first
fixing a finite-dimensional vector space of one-electron wave functions (the \emph{spin-orbital space}):
\begin{equation}
  \label{eq:orb_space}
  \orbSp = \text{span}\{ \phi_p \}_{p=1}^M\,.
\end{equation}
This $M$-dimensional vector space is usually defined by
choosing a basis set for an electronic structure calculation.
From the spin-orbital space
one can construct $n$-electron wave functions as linear combinations of $n$-electron Slater determinants made by elements of $\orbSp$.
These $n$-electron wave functions form the required vector space, that is the $n$-th \emph{exterior power} of $\orbSp$, denoted by $\extProd{}$ \cite{ruiz-tolosa-05_0_0}.
From the point of view of the exterior algebra, an $n$-electron Slater determinant is the \emph{exterior product}, or \emph{wedge product}, $\w$, of $n$ elements of $\orbSp$.
For instance:
\begin{equation}\label{eq:arbitrary_decomposable}
  \ket{\Phi} = 
  \frac{1}{\sqrt{n!}}
  \begin{vmatrix}
    \varphi_1(1) & \varphi_2(1) & \dots & \varphi_n(1)\\ 
    \varphi_1(2) & \varphi_2(2) & \dots & \varphi_n(2)\\
    \vdots & \vdots & \ddots & \vdots\\
    \varphi_1(n) & \varphi_2(n) & \dots & \varphi_n(n)\\ 
  \end{vmatrix}
  = \varphi_1 \w \varphi_2 \w \dots \w \varphi_n\,,
\end{equation}
where $\{\varphi_i\}_{i=1}^n$ is a linearly independent set, but otherwise arbitrary, of elements of $\orbSp$.
Therefore the vector space $\extProd{}$ consists of all possible linear combinations of such elements,
and it is identified with the $n$-electron sector of the Fock space
(whereas the complete Fock space is identified with the \emph{exterior algebra} of $\orbSp$, $\extAlg =\bigoplus_{n=0}^M \extProd$).
In this article, \emph{ket} notation will be used for $n$-electron wave functions,
whereas small Greek letters are used for one-electron wave functions (orbitals).
Besides subscripts, Slater determinants will be generally denoted by $\ket{\Phi}$
and arbitrary $n$-electron wave functions by $\ket{\Psi}$.


Elements such as $\varphi_1 \w \dots \w \varphi_n \in \extProd{}$, that can be written as the exterior product of elements of $\orbSp{}$, are said to be \emph{decomposable}
(the nomenclatures \emph{simple} and \emph{free} are also used by some authors,
the latter especially in the context of quantum entanglement \cite{dingJCTC20_16_4159}).
Slater determinants are thus the decomposable elements of $\extProd$.
Furthermore, given a basis for $\orbSp$, such as in Eq.~\eqref{eq:orb_space}, the set of all decomposable elements made by $n$ elements of this basis (with no repetition and taken in order, e.g by ascending indices), forms a basis for $\extProd$.
An arbitrary element of $\extProd{}$ can be constructed as:
\begin{equation}\label{eq:gen_elem_ext_alg}
  \begin{split}
    \ket{\Psi} &= \sum_{I,\, |I|=n} C_I \, \phi_{I_1} \w \dots \w \phi_{I_n}\\
     &= \sum_{I,\, |I|=n} C_I \, a_{I_1}^\dagger \dots a_{I_n}^\dagger \ket{}\\
     &= \sum_{I,\, |I|=n} C_I \ket{\Phi_I}\,,
  \end{split}
\end{equation}
where the summation runs over all multi-indices sets $I$ with length $n$.
For convenience, the indication $|I| = n$ will be often dropped,
as the length of the multi-indices set is always the number of electrons,
that will be clear by the context.
Eq.~\eqref{eq:gen_elem_ext_alg} is clearly interpreted as a configuration interaction (CI) expansion.
We also make the connection with the formalism of second quantization, with $a^\dagger$ being creation operators and $\ket{}$ the vacuum state.
Note that $\ket{\Psi}$ might be decomposable or not, depending whether there is a basis for $\orbSp{}$, say $\{\phi'_p\}_{p=1}^M$, such that 
\begin{equation}
  \ket{\Psi} = \phi'_1 \w \dots \w \phi'_n\,.
\end{equation}
In general, given a wave function in the form of Eq.~\eqref{eq:gen_elem_ext_alg},
it is not evident if it is decomposable.
We want to find a characterization of the set of all wave functions in $\extProd$ that are decomposable.



Consider the following Slater determinant:
\begin{equation}\label{eq:slater_det_gen}
  \ket{\Phi} = \varphi_1 \w \dots \w \varphi_n\,,
\end{equation}
with each $\varphi_i$ being an element of $\orbSp{}$.
If the set $\{\varphi_1, \dots, \varphi_n\}$ is linearly dependent, the Slater determinant vanishes, namely, it is the zero element of the vector space $\extProd{}$.
Assuming that this set is linearly independent, it spans a $n$-dimensional vector subspace of $\orbSp$:
\begin{equation}
  \text{span}\{ \varphi_1, \dots, \varphi_n \} \subset \orbSp{}
\end{equation}
\begin{equation}
  \text{dim}(\text{span}\{ \varphi_1, \dots, \varphi_n \}) = n
\end{equation}
Obviously, this vector space admits infinitely many other basis, obtained from $\{\varphi_1, \dots, \varphi_n\}$ by a non singular linear transformation:
\begin{equation}
  \varphi_i' = \sum_{p=1}^M \varphi_p U_i^p
\end{equation}
\begin{equation}\label{eq:two_basis_for_Slater_det}
  \text{span}\{ \varphi_1, \dots, \varphi_n \} = \text{span}\{ \varphi'_1, \dots, \varphi'_n \}\,.
\end{equation}
Furthermore, and of central importance for the present argument, the decomposable element made by $\varphi'_i$ differs to $\ket{\Phi}$ (Eq.~\eqref{eq:slater_det_gen}) by normalization or phase only:
\begin{equation}
  \varphi'_1 \w \dots \w \varphi'_n = \lambda \varphi_1 \w \dots \w \varphi_n\,
  \quad\lambda \ne 0\,.
\end{equation}
Thus, there is a one-to-one map between the physical states that can be represented by a Slater determinant and the set of $n$-dimensional subspaces of $\orbSp$.
The set of all $n$-dimensional vector subspaces of a given vector space $\orbSp$ is the \emph{Grassmannian}, or the \emph{Grassmann manifold}.
It will be represented by $\grass{n}{\orbSp}$.


Recall that the normalization and phase of a wave function is not relevant for the description of the physical state it represents.
Thus, the space of states for the $n$-electron system is actually the \emph{projective space} of $\extProd$, denoted by $\projective \extProd$.
This space is the set of the equivalence classes in $\extProd{}$ obtained by the relation $\ket{\Psi} \sim \lambda \ket{\Psi}$, where $\lambda$ is a nonzero scalar.
The equivalence class of $\ket{\Psi}$, indicated by $[\ket{\Psi}]$, is composed by the wave functions that differ from $\ket{\Psi}$ by a normalization or phase factor.


From the above discussion,
an element of $\grass{n}{\orbSp}$ (a vector subspace of $\orbSp$) is associated to the equivalence class of a Slater determinant (an element of $\projective\extProd$).
This is a bijection with the set of all Slater determinants (except for a scalar factor),
that forms a submanifold in the space of wave functions that is a copy of the Grassmannian $\grass{n}{\orbSp}$ inside $\projective\extProd$:
\begin{eqnarray}
  \text{Slater determinants} & \hookrightarrow & \text{$n$-electron wave functions}\quad\\
  \grass{n}{\orbSp{}} &\hookrightarrow& \projective \extProd{}\\\label{eq:plucker}
  {[}\ket{\Phi}] &\mapsto& [\varphi_1 \w \dots \w \varphi_n]\,.
\end{eqnarray}
The application given in Eq.~\eqref{eq:plucker} is known as the \emph{Pl{\"u}cker embedding} \cite{griffiths-78_0_0,hodge-94_2_0}.
The image of $\grass{n}{\orbSp{}}$ in $\projective \extProd{}$ satisfies a set of quadratic equations in $\projective\extProd$,
the \emph{Pl{\"u}cker relations},
of high importance in the field of algebraic and projective geometry.
Thus, an element $\ket{\Psi} \in \extProd$ is decomposable if and only if its coefficients on a basis made by decomposable elements
(as in Eq.~\eqref{eq:gen_elem_ext_alg}) satisfy the Pl{\"u}cker relations.
A particular case of these relations will be shown in Sec.~\ref{sec:ex_hydrogen}.


The strong connection between Slater determinants and the Grassmannian suggests that the properties of the latter can be used to work with the former, in particular for their optimization.
This will be explored in the remaining of this article.
Thus, we will make no distinction between a wave function, that is a point in $\extProd$,
and its equivalence class in $\projective \extProd$;
we will often say that a wave function is at the Grassmannian,
or belongs to the Grassmannian, when it can be represented by a Slater determinant;
we will also interchange the nomenclatures, and make no distinction between decomposable elements of $\extProd$, Slater determinants, and the vector subspace of $\orbSp$ spanned by its orbitals:
\begin{eqnarray}\label{eq:association_vec_sp_Slater_det}
  [\ket{\Phi}] =&& [\varphi_1 \w \dots \w \varphi_n]\nonumber\\
  =&& \text{span}\{ \varphi_1, \dots, \varphi_n \} \in \grass{n}{\orbSp} \subset \projective \extProd\,.
\end{eqnarray}

\subsection{\label{sec:repr_grass}Representation of the Grassmannian}


The most obvious way to represent a Slater determinant is by a $M \times n$ matrix of rank $n$,
denoted by $U$,
having the coefficients of a basis of the Slater determinant in a fixed basis of $\orbSp$
(e.g., of Eq.~\eqref{eq:orb_space}):
\begin{equation}\label{eq:def_matrix_U}
  \ket{\Phi} = \phi'_1 \w \dots \w \phi'_n \quad \Leftrightarrow \quad U \in \real^{M\times n}
\end{equation}
\begin{equation}\label{eq:def_matrix_U_entries}
  \phi'_i = \sum_{p=1}^M \phi_pU_i^p\,.
\end{equation}
Thus, column $i$ has the $M$ coefficients of orbital $\phi'_i$ on this basis of $\orbSp$.
Because $U$ is of rank $n$,
its columns are linearly independent and span an $n$-dimensional vector space,
associated to the $n$-electron Slater determinant.
We write $[\ket{\Phi}] = \text{span}(U)$.
This matrix is not unique, see Eq.~\eqref{eq:two_basis_for_Slater_det}.
Thus,
to run over all the Grassmannian when looking for an optimal Slater determinant of any sort,
the entries of $U$ cannot be freely varied,
because a change on the entries of $U$ might provoke no change on the corresponding Slater determinant,
or lead to a matrix $U$ with linearly dependent columns (that does not span an $n$-dimensional vector space and does not represent an $n$-electron Slater determinant).


Slater determinants can also be parametrized starting from a pivot Slater determinant, say
\begin{equation}\label{eq:def_phi_ini}
  \ket{\Phi_0} = \phi_1 \w \dots \w \phi_n\,,
\end{equation}
by
\cite{thoulessNP60_21_225,rowePRA80_22_2363,helgaker00_molec}:
\begin{equation}\label{eq:orb_rot_exp}
  \ket{\Phi} = e^{\hat{K}}\ket{\Phi_0}\,,
\end{equation}
where
\begin{equation}\label{eq:def_orb_rot_parameters}
  \hat{K} = \sum_{i=1}^n\sum_{a=n+1}^M K_i^a \left( a_a^\dagger a_i - a_i^\dagger a_a \right)\,.
\end{equation}
A transformation matrix from this basis to a basis of $\ket{\Phi}$ is given by:
\begin{equation}
  \label{eq:def_new_basis_orb_rot}
  U_{\text{full}} = \exp\left\{
      \begin{pmatrix}
        \mathbf{0}_{n \times n} & \mathbf{-K}^T\\
        \mathbf{K} & \mathbf{0}_{(M-n) \times (M-n)}
      \end{pmatrix}
    \right\}\,.
\end{equation}
Note that the matrix $U$ in Eq.~\eqref{eq:def_matrix_U} corresponds to the first $n$ columns of $U_{\text{full}}$.


This second parametrization is originated from the works of Thouless \cite{thoulessNP60_21_225},
and has been studied by several authors in the context of electronic structure theory
\cite{linderbergIJQC77_XII_161,dalgaardJCP78_69_3833,yeagerJCP79_71_755,rowePRA80_22_2363}.
It offers a set of $n(M-n)$ \emph{independent} parameters $K_i^a$ (not considering possible symmetry constraints),
that is exactly the dimension of the Grassmannian \cite{borisenkoRMS91_46_45}.
This parametrization is largely used to carry out variations on the orbitals in SCF methods
\cite{siegbahnPS80_21_323,siegbahnJCP81_74_2384,wernerJCP85_62_5053,werner-87_0_0,shepard-87_0_0,roos-87_0_0,helgaker00_molec},
and one often refers to occupied-virtual (and other ``type X-type Y'') orbital rotations in orbital optimizations.
It is the \emph{de facto} parametrization used in modern SCF calculations.
However, in Sec.~\ref{sec:alg_Absil} we will show that the first type of representation described above,
Eq.~\eqref{eq:def_matrix_U},
can also be used for a Newton optimization of Slater determinants,
contrary to what is often assumed \cite{werner-87_0_0,shepard-87_0_0,helgaker00_molec}.

\subsection{\label{sec:ex_hydrogen}Example: the hydrogen molecule}

\begin{figure}
  \includegraphics[width=0.5\textwidth]{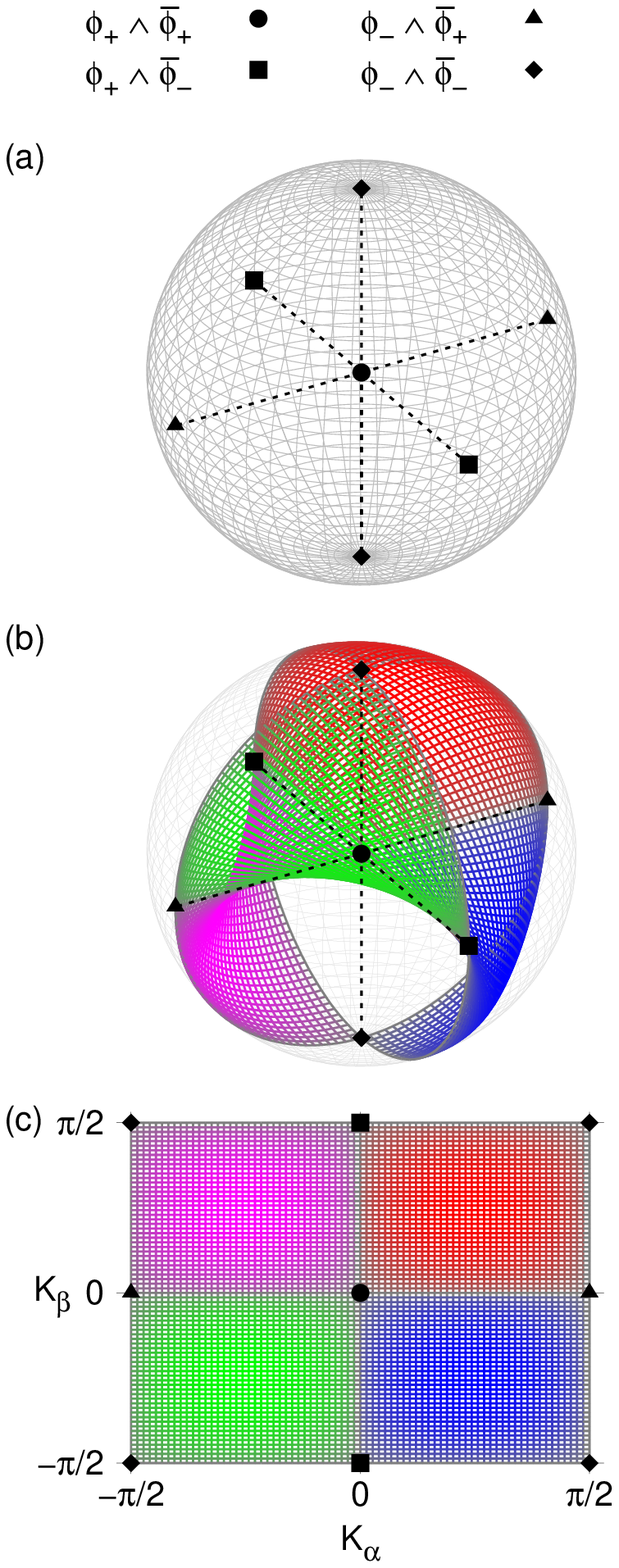}
  \caption{\label{fig:proj_space_H2}(a) A representation of $\projective \big(\extProdTwo{}_{H_2}\big)_{M_S=0}$,
    along with the four Slater determinants given in Eq.~\eqref{eq:wf_space_H2_Ms0};
    (b) The $M_S = 0$ subset of the Grassmannian $\grass{2}{\orbSp_{H_2}}$ embedded in $\projective \big(\extProdTwo{}_{H_2}\big)_{M_S=0}$;
    (c) The space of the parameters $K_\alpha$ and $K_\beta$, that determines the surface at b) by Eq.~\eqref{eq:general_Phi_H2}.
    }
\end{figure}

We will illustrate the concepts discussed so far for the $M_S = 0$ states of the hydrogen molecule, H$_2$,
described by a minimal basis set
($1s$ alpha and beta orbitals centered in each atom),
with real coefficients and orbitals.
This case allows a visualization of the Grassmannian as embedded in the projective space of the two-electron wave functions,
depicted in Fig.~\ref{fig:proj_space_H2}.
For convenience, we will use symmetry adapted orbitals:
\begin{eqnarray}
  \phi_+ &=& N_+(1s_A + 1s_B)\\
  \phi_- &=& N_-(1s_A - 1s_B)\,,
\end{eqnarray}
with $N_\pm = \frac{1}{\sqrt{2 \big( 1 \pm \bracketbig{1s_A}{1s_B}\big)}}$,
but the present discussion does not depend by any means on this particular basis,
and only in Sec.~\ref{sec:opt_ex_H2} it will become apparent the reason for this choice.
The orbital and two-electron wave function spaces are (beta-spin orbitals are indicated by over-lines):
\begin{equation}
  \label{eq:orb_space_H2}
  \orbSp_{H_2} = \text{span}\{
  \phi_+, \phi_-,
  \overline{\phi_+}, \overline{\phi_-} \}
\end{equation}
\begin{equation}
  \label{eq:wf_space_H2}
  \begin{split}
    \extProdTwo{}_{H_2} = \text{span}\{&
    \phi_+ \w \overline{\phi_+},
    \phi_+ \w \overline{\phi_-},\\
    &\phi_- \w \overline{\phi_+},
    \phi_- \w \overline{\phi_-},\\
    &\phi_+ \w \phi_-,
    \overline{\phi}_+ \w \overline{\phi}_-\}
  \end{split}
\end{equation}
\begin{equation}
  \label{eq:wf_space_H2_Ms0}
  \begin{split}
    \Big(\extProdTwo{}_{H_2}\Big)_{M_S=0} = \text{span}\{
    &\phi_+ \w \overline{\phi_+},
    \phi_+ \w \overline{\phi_-},\\
    &\phi_- \w \overline{\phi_+},
    \phi_- \w \overline{\phi_-}\}\,.
  \end{split}
\end{equation}
As $\big(\extProdTwo{}_{H_2}\big)_{M_S=0}$ is four-dimensional its projective space is three-dimensional.
An arbitrary element of $\big(\extProdTwo{}_{H_2}\big)_{M_S=0}$ is:
\begin{equation}\label{eq:basis_Ms0_H2}
  \ket{\Psi} = C_{13} \, \phi_+ \w \overline{\phi_+}
  + C_{14}\, \phi_+ \w \overline{\phi_-}
  + C_{23}\, \phi_- \w \overline{\phi_+}
  + C_{24}\, \phi_- \w \overline{\phi_-}\,.
\end{equation}
A visual representation of this space can be obtained in the following way:
The elements of $\projective \big(\extProdTwo{}_{H_2}\big)_{M_S=0}$ can be viewed as the ``rays''
(straight lines that pass through the origin)
in a four-dimensional space.
Each of these lines cross the unit sphere $S^3 \subset \real{}^4$ in two antipodal points.
They represent the two normalized wave functions in $\big( \extProdTwo{}_{H_2}\big)_{M_S=0}$
associated to the same physical state, but differing by sign.
To fix one representative element, we choose the one with non negative coordinate for (say)
$\phi_+ \w \overline{\phi_+}$.
That is, the ``upper hemisphere'' of $S^3$ with respect to the direction of $\phi_+ \w \overline{\phi_+}$.
Finally, we project these points of $S^3 \subset \real^4$ into the $\real^3$ space defined by $C_{13} = 0$,
obtaining the unit ball in $\real^3$,
so that each of its points represents an element of $\projective \big(\extProdTwo{}_{H_2}\big)_{M_S=0}$.
This representation is depicted in Fig.~\ref{fig:proj_space_H2}.a).
It can be interpreted as what an observer in $\real^4$ would see,
when looking at $S^3$ from the top.
At the center of the visualization plane there is $[\phi_+ \w \overline{\phi_+}]$
(as this observer is looking exactly from its direction),
and the other elements of the basis in Eq.~\eqref{eq:basis_Ms0_H2} are represented (twice) in the boundary of the ball.

The Pl{\"u}cker relation that characterizes how the Grassmannian $\grass{2}{\orbSp_{H_2}}$ is embedded in $\projective \big(\extProdTwo{}_{H_2}\big)$ is
(there are much more equations, with more terms, for larger cases) \cite{griffiths-78_0_0}:
\begin{equation}\label{eq:Plucker}
  C_{12}C_{34} - C_{13}C_{24} + C_{14}C_{23} = 0\,,
\end{equation}
where $C_{12}$ and $C_{34}$ are the coefficients of Slater determinants with $M_S \ne 0$,
and are zero in the present case.
Furthermore, we are choosing normalized wave functions with $C_{13} \ge 0$,
and thus the relation becomes:
\begin{equation}\label{eq:Plucker_specific}
  C_{24}\sqrt{1 - (C_{24}^2 + C_{14}^2 + C_{23}^2)} - C_{14}C_{23} = 0\,.
\end{equation}
The Grassmannian is represented by the set of points in the space of variables $\{C_{24}, C_{14}, C_{23}\}$ that satisfies this relation,
and the corresponding wave functions can be written as single Slater determinants.
Fig.~\ref{fig:proj_space_H2}.b) illustrates how this manifold is embedded in $\projective \big(\extProdTwo{}_{H_2}\big)$,
with the representation described above.

Observe that the straight lines seen in Fig.~\ref{fig:proj_space_H2}.a) and ~\ref{fig:proj_space_H2}.b),
joining $\phi_+ \w \overline{\phi_+}$ with each one of the other Slater determinants in the basis of Eq.~\eqref{eq:basis_Ms0_H2},
represent wave functions that are linear combinations of $\phi_+ \w \overline{\phi_+}$ with only one other element of Eq.~\eqref{eq:basis_Ms0_H2}.
For instance, the vertical line represents wave functions of the form:
\begin{equation}\label{eq:doubly_exc_lin_comb_H2}
  \ket{\Psi} = C_{13} \, \phi_+ \w \overline{\phi_+}
  + C_{24}\, \phi_- \w \overline{\phi_-}\,.
\end{equation}
In the ``north hemisphere'', $C_{13}$ and $C_{24}$ have the same sign,
whereas in the ``south hemisphere'' they have opposite signs.
Except for the cases when one of $C_{13}$ or $C_{24}$ is zero,
the points of this line are not at the Grassmannian, see Fig.~\ref{fig:proj_space_H2}.b),
what means that the wave function above cannot in general be taken as a single Slater determinant,
a well known fact in electronic structure.
On the other hand, the horizontal straight lines, connecting $\phi_+ \w \overline{\phi_+}$ and $\phi_+ \w \overline{\phi_-}$ or $\phi_+ \w \overline{\phi_-}$, are at the Grassmannian.
The corresponding wave functions are
\begin{equation}
  \begin{split}
    \ket{\Phi_1} =& C_{13} \, \phi_+ \w \overline{\phi_+}
    + C_{23}\, \phi_- \w \overline{\phi_+}\\
    \ket{\Phi_2} =& C_{13} \, \phi_+ \w \overline{\phi_+}
    + C_{14}\, \phi_+ \w \overline{\phi_-}\,,
  \end{split}
\end{equation}
that are linear combinations of $\phi_+ \w \overline{\phi_+}$ with single excitations from $\phi_+$ to $\phi_-$ (at alpha or beta spin, respectively).
As it turns out to be, linear combinations with singly excited determinants can always be represented as a single Slater determinant:
\begin{equation}\label{eq:singly_exc_lin_comb_H2_version2}
  \begin{split}
    \ket{\Phi_1} =& (C_{13} \, \phi_+  + C_{23}\, \phi_-) \w \overline{\phi_+}\\
    \ket{\Phi_2} =& \phi_+ \w (C_{13} \, \overline{\phi_+} + C_{14}\, \overline{\phi_-})\,.
  \end{split}
\end{equation}
The fact that $\ket{\Psi}$ (in Eq.~\eqref{eq:doubly_exc_lin_comb_H2}) does not belong to the Grassmannian in general, whereas $\ket{\Phi_1}$ and $\ket{\Phi_2}$ do, is clearly seen in the Pl{\"u}cker relations, Eq.~\eqref{eq:Plucker} and \eqref{eq:Plucker_specific}:
the coefficients of $\ket{\Psi}$ do not satisfy Eq.~\eqref{eq:Plucker_specific},
but the coefficients of $\ket{\Phi_1}$ and $\ket{\Phi_2}$ do.

Equation~\eqref{eq:singly_exc_lin_comb_H2_version2} suggests that every point of the $M_S = 0$ subset
of the Grassmannian $\grass{2}{\orbSp_{H_2}}$ can be obtained as
\begin{equation}
  \ket{\Phi} =
  \left( a \, \phi_+
    + b\, \phi_- \right)
  \w \left( c \, \overline{\phi_+}
    + d \, \overline{\phi_-} \right)\,,
\end{equation}
to be represented by the matrix
\begin{equation}\label{eq:param_U_H2}
  U =
  \begin{pmatrix}
    a & 0 \\
    b & 0 \\
    0 & c\\
    0 & d
  \end{pmatrix}\,.
\end{equation}
The division into blocks comes from the fact that we are concerned with a single value for $M_S$, what naturally excludes mixing among alpha and beta orbitals.
On the other hand,
the orbital $a \, \phi_+ + b \, \phi_-$ can also be interpreted as a rotation of $\phi_+$ towards $\phi_-$
by an angle of ${K_\alpha} = \arctan(\frac{b}{a})$.
Analogously, one defines ${K_\beta} = \arctan(\frac{d}{c})$.
The corresponding normalized Slater determinant, made of normalized orbitals, becomes:
 \begin{eqnarray}\label{eq:general_Phi_H2}
  \ket{\Phi} =&&
                 \left( \cos ({K_\alpha}) \, \phi_+
                 + \sin ({K_\alpha}) \, \phi_- \right)\nonumber\\
                 &&\w \left( \cos ({K_\beta}) \, \overline{\phi_+}
                 + \sin ({K_\beta}) \, \overline{\phi_-} \right)\nonumber\\
             =&&  e^{\hat{K}} \big( \phi_+ \w \overline{\phi_+} \big)\,,
 \end{eqnarray}
 where $\hat{K}$ is defined in Eq.~\eqref{eq:def_orb_rot_parameters}.
 Equations~\eqref{eq:param_U_H2} and \eqref{eq:general_Phi_H2} are the two parametrizations of the Grassmannian, based on the coefficients matrices and on orbital rotations, described in Sec.~\ref{sec:repr_grass}.
 Finally,
 the $M_S = 0$ subset of the Grassmannian is complete if the space of the parameters ${K_\alpha}$ and ${K_\beta}$ is
 $[-\frac{\pi}{2}, \frac{\pi}{2}] \times [-\frac{\pi}{2}, \frac{\pi}{2}]$,
 as shown in Fig.~\ref{fig:proj_space_H2}.c).

 An analogous representation of the $M_S = 0$ subset of the Grassmannian
 has been obtained by Cassam-Chena{\"i} \cite{cassam-chenaiJMC94_15_303},
 although he represents the projective space in a more pictorial fashion.
 The reader is strongly referred to his work for a deeper discussion of the concepts presented in this section.

\subsection{\label{sec:the_problem}The overlap with a correlated wave function}


Consider now an arbitrary wave function, $\ket{\extWF} \in \extProd$,
that does not necessarily belong to the Grassmannian
(subscript ``ext'' stands for ``external to the Grassmannian'').
For instance, this could be the exact or some approximate wave function for the ground state of the system.
The objective of this work is to devise and study algorithms to find the Slater determinant that maximizes the overlap to this wave function.
This makes sense only for the equivalence classes of $\ket{\extWF}$ and of the Slater determinants, their corresponding elements in $\projective \extProd$.
Thus, our problem consists in optimizing the following function, defined at the Grassmannian:
\begin{equation}\label{eq:the_f}
  f(\left[ \ket{\Phi} \right]) = \frac{|\bracket{\Phi}{\extWF}|}
  {\sqrt{{\bracket{\Phi}{\Phi}
        \bracket{\extWF}{\extWF}}}}
\end{equation}

\section{\label{sec:algorithms}Algorithms}


In this section we will discuss algorithms for the optimization of the function defined in Eq.~\eqref{eq:the_f}, using the Newton method.
To simplify the analysis and the discussion of the equations, they will be presented on a spin-orbital basis, with no inclusion of spatial symmetry.
Complete equations using spatial orbitals and considering spatial symmetry from Abelian point groups are given in the Appendix.

\subsection{\label{sec:alg_orb_rot}Algorithm 1: by
  independent parameters in
  $\ket{\Phi} = e^{\hat{K}} \ket{\Phi_0}$}


With the parametrization given by Eq.~\eqref{eq:orb_rot_exp} one can directly apply the Newton method.
Let 
\begin{eqnarray}
  \nonumber f_{\square} : \real^{n(M - n)} &\to& \real\\
  \Kmat &\mapsto& f(e^{\hat{K}}\ket{\Phi_0})\,,
\end{eqnarray}
be the function that represents the overlap function $f$ (Eq.~\eqref{eq:the_f}),
but defined over the space of orbital rotation parameters.
An improved Slater determinant is obtained from $\ket{\Phi_0}$ with the parameters $K_i^a$ (collected in $\Kmat$) that solve the equation
\begin{equation}\label{eq:orb_rot_Newton_step}
  \Hess{} \Kmat = -\Jac{}\,,
\end{equation}
where $\Jac{}$ and $\Hess$, the Jacobian and Hessian of $f_\square$, are made by the first and second derivatives of $f_\square$.
Calculation of $f_\square$, $\Jac$ and $\Hess$ is straightforward when $\Kmat=0$,
and $\ket{\extWF}$ is given as a normalized linear combination of excitations on top of $\ket{\Phi_0}$:
\begin{equation}\label{eq:psi_ext}
  \ket{\extWF} =
  C_0 \ket{\Phi_0}
  + \sum_{i,a} C_i^a \ket{\Phi_i^a}
  + \frac{1}{4}\sum_{i,j,a,b} C_{ij}^{ab} \ket{\Phi_{ij}^{ab}}
  + \dots\,,
\end{equation}
with $C_{ij}^{ab} = C_{ji}^{ba} = -C_{ij}^{ba} = -C_{ji}^{ab}$.
In such conditions:
\begin{equation}\label{eq:the_f_orb_rot}
  f_\square(\Kmat = 0) = C_0
\end{equation}
\begin{equation}\label{eq:jac_alg_orb_rot}
  \Jac_i^a = \frac{\partial f_\square(\Kmat = 0)}{\partial K_i^a}
  = (-1)^{i + n} C_i^a
\end{equation}
\begin{equation}\label{eq:hess_alg_orb_rot}
  \Hess_{ij}^{ab} =
  \frac{\partial^2 f_\square(\Kmat = 0)}{\partial K_i^a \partial K_j^b} =
  \left\{
    \begin{array}{lcr}
      - C_0 & \quad& a=b, i=j\\
      -(-1)^{i+j} C_{ij}^{ab} & \quad& \text{otherwise}
    \end{array}
  \right.\,.
\end{equation}


Note that the Jacobian is constructed from the coefficients of single excitations,
whereas the Hessian is formed by the coefficients of the reference (diagonal elements) and of the double excitations (all with respect to $\ket{\Phi_0}$).
However, these expressions are valid only if $\Kmat=0$, otherwise higher rank excitations also contribute, in a non trivial way.
Hence, if this algorithm is used to optimize $f$,
the basis of $\orbSp$ used to expand $\ket{\extWF}$ has to be changed in every iteration,
to obtain the coefficients as excitations with respect to the new Slater determinant $\ket{\Phi} = e^{\hat{K}}\ket{\Phi_0}$.


Changing the orbital basis used to expand $\ket{\extWF}$ is very disadvantageous.
First of all because this is a time consuming step.
If carried out in a straightforward way, it is accomplished by the formula:
\begin{eqnarray}\label{eq:FCI_transf}
  \ket{\extWF} &=& \sum_I C_I \ket{\Phi_I}
                            = \sum_I C_I \, \phi_{I_1} \w \dots \w \phi_{I_n}\nonumber\\
                        &=& \sum_I C'_I \ket{\Phi'_I}
                            = \sum_I C'_I \, \phi'_{I_1} \w \dots \w \phi'_{I_n}
\end{eqnarray}
\begin{equation}\label{eq:trans_fci_orbital_basis}
  C'_I = \sum_J C_J\, \text{det}\big((U_\text{full})^I_J\big)\,,
\end{equation}
where $U_\text{full}$ is the transformation matrix from the basis $\{\phi'_p\}$ to the basis $\{\phi_p\}$, obtained from $\Kmat$ by Eq.~\eqref{eq:def_new_basis_orb_rot}.
The matrices $(U_\text{full})^I_J$ are the minors of the matrix $U_\text{full}$, with the entries in the rows and columns given by the multi-indices $I$ and $J$.
This is the bottleneck step, and its computational cost is discussed in Sect.~\ref{sec:comp_cost}.
However, the main disadvantage is that the representation of the external wave function is changed in every iteration.
This implies that,
if $\ket{\extWF}$ is an approximate wave function based on some kind of rank truncation,
say, a configuration interaction with single and double excitations
(CISD) wave function over the reference $\ket{\Phi_0}$,
it will not contain only single and double excitations over the new Slater determinant $\ket{\Phi} = e^{\hat{K}} \ket{\Phi_0}$.
Therefore, the initial \emph{structure} of this wave function is lost,
and a full configuration interaction (FCI)-like wave function has always to be used as external wave function.
Throughout this text,
any possible rank truncation scheme used to construct $\ket{\extWF}$ will be generally denoted as the structure of $\ket{\extWF}$.

\subsection{\label{sec:alg_Absil}Algorithm 2:
  using the coefficients matrix $U$}


The difficulties in the algorithm above arise from the standard
parametrization by orbital rotations, Eq.~\eqref{eq:orb_rot_exp}.
It is often assumed that a set of independent parameters is necessary to perform a Newton optimization of orbitals,
as otherwise the Hessian matrix is singular or near-singular close to the optimal orbitals \cite{helgaker00_molec}.
This would exclude the possibility of using directly the matrix $U$, defined in Eq.~\eqref{eq:def_matrix_U}, for such kind of optimization.
However, this is exactly what the optimization procedure on the Grassmannian of Absil and coworkers do \cite{absilAAM04_80_199}.


Using the $M \times n$ matrices of full rank to represent the elements of the Grassmannian,
Absil and coworkers have studied the Riemannian geometry of the Grassmannian.
These matrices form the non-compact Stiefel manifold, ST($n$, $M$),
and all computations are carried out on it,
whereas tools of differential geometry are used to go back and forth from the Stiefel manifold to the Grassmann manifold.
With this technique, Absil and coworkers presented several formulas for geometric concepts on Grassmannians,
such as canonical metric and geodesics.
More important for the present work, a Newton method specific for the Grassmann manifold was also proposed.
It works in the following way \cite{absilAAM04_80_199}:
Given $f : \grass{n}{\orbSp{}} \to \real$ a function defined on the Grassmannian with real values sufficiently smooth, let
\begin{eqnarray}
  \nonumber f_{\lozenge} : \stiefel{n}{M} &\to& \real\\
  U &\mapsto& f(\text{span}(U))\,,
\end{eqnarray}
be the corresponding function on the Stiefel manifold, that is,
the function defined over $M \times n$ matrices that,
when calculated on any representative matrix $U$ of $\ket{\Phi_0}$,
returns $f([\ket{\Phi_0}])$.
The procedure is carried out by computations over the matrix $U$,
that belongs to the Stiefel manifold:
\begin{itemize}
\item One first solves the following equation for the unknown $\eta_{\lozenge U} \in H_U = \{U_\perp K:K \in \real^{(M-n)\times n}\}$:
  \begin{equation}\label{eq:Absil_main_eq}
    \Pi_{U_\perp} D\left(\Pi_{\cdot{}_\perp} \text{grad}f_\lozenge\left(\cdot{}\right) \right)
    \left(U\right)\left[\eta_{\lozenge U}\right] = -\Pi_{U_\perp}\text{grad}f_\lozenge\left(U\right)\,;
  \end{equation}
\item And update $\ket{\Phi_0} \to \ket{\Phi}$ by moving along the geodesic on the Grassmannian in the direction of $\eta_{\lozenge U}$,
  by computing a singular value decomposition (SVD) of
  $\eta_{\lozenge U} = \mathcal{U} \Sigma \mathcal{V}^T$ and calculating:
  \begin{equation}\label{eq:Absil_SVD}
    [\ket{\Phi}] = \text{span}(U \mathcal{V} \cos \Sigma + \mathcal{U} \sin \Sigma)\,.
  \end{equation}
\end{itemize}
In these equations,
$U_\perp$ is any full-rank $M \times (M-n)$ matrix such that $U^T U_\perp = 0$;
the gradient of $f_\lozenge$ at $U$ is the $M \times n$ matrix whose entries are given by
$(\text{grad} f_\lozenge(U))_q^p = \frac{\partial f_\lozenge(U)}{\partial U_q^p}(U)$;
$DF(x)[y] = \frac{d}{dt}F(x + ty)\big|_{t=0}$
is the directional derivative of $F$ at $x$ in the direction of $y$;
$\Pi_{W_\perp} = I - W(W^TW)^{-1}W^T$
is the projection onto the orthogonal complement of the matrix $W$;
The dot $\cdot$ in Eq.~\ref{eq:Absil_main_eq} denotes the point where the function has to be evaluated
(and thus it stands for $U+t\eta_{\lozenge U}$ when calculating the directional derivative);
and the matrices $\mathcal{U}$, $\Sigma$, and $\mathcal{V}$ are $M \times n$ orthonormal,
$n \times n$ diagonal, and $n \times n$ orthonormal, respectively.


There is a clear analogy between Eq.~\eqref{eq:Absil_main_eq} and the standard Newton method,
Eq.~\eqref{eq:orb_rot_Newton_step}:
the left-hand side of Eq.~\eqref{eq:Absil_main_eq} is related to the second derivatives of $f_\lozenge$,
calculated in the direction of $\eta_{\lozenge U}$,
whereas in the right-hand side are the first derivatives.
However, the projectors onto the orthogonal complement of $U$
and the requirement that the unknown $\eta_{\lozenge U}$ belongs to $H_U$
(for every element $\eta$ of $H_U$, $U^T\eta = 0$)
guarantee that variations in $U$ that do not change $\ket{\Phi_0}$ are canceled out
\cite{absilAAM04_80_199}.
Thus, the usage of a redundant set of parameters does not pose a problem here.


We have adapted this procedure for the overlap function $f$, defined in Eq.~\eqref{eq:the_f}.
Assuming $U$ orthogonal, $U^T U = \mathbb{1}$, this function can be calculated as
\begin{equation}\label{eq:our_f}
 f([\ket{\Phi}]) = f_\lozenge(U)
  =  \sum_{I} C_I \, F_I\,,
\end{equation}
and Eq.~\eqref{eq:Absil_main_eq} becomes the following system of linear equations:
\begin{equation}
  \label{eq:gen_lin_system_Absil}
  \sum_{r=1}^M\sum_{s=1}^n \Habsil_{qs}^{pr} \big(\eta_{\lozenge U}\big)^r_s = - \Jabsil_q^p
\end{equation}
where:
\begin{eqnarray}
  \label{eq:def_Habsil}
  \Habsil_{qs}^{pr} &=&  \sum_{\bar{p} = 1}^M
                        \big( \Pi_{U_\perp} \big)_{\bar{p}}^p
                        \sum_I C_I
                        \big( \mathbf{\tilde{H}}_I \big)_{qs}^{{\bar{p}} r}\\
  \label{eq:def_Jabsil}
  \Jabsil_q^p &=&
                  \sum_{\bar{p} = 1}^M
                  \big( \Pi_{U_\perp} \big)_{\bar{p}}^p
                  \sum_I C_I  \big( \matG_I \big)_q^{\bar{p}}\\
  \label{eq:def_FI}
  \big( F_I \big) &=& \text{det} \left( U \big|_I\right) \\
  \label{eq:def_GI}
  \big( \matG_I \big)_q^p &=& \text{det} \left( (U \overset{q}{\leftarrow} e_p )\big|_I \right)\\
  \label{eq:def_HI}
  \big( \mathbf{H}_I \big)_{qs}^{pr} &=& \text{det} \left( (U \overset{q}{\leftarrow} e_p
                                \overset{s}{\leftarrow} e_r )\big|_I \right)\\
  \label{eq:def_HI_tilde}
  \big( \mathbf{\tilde{H}}_I \big)_{qs}^{pr} &=& \left\{
                                        \begin{array}{ll}
                                          \big( \mathbf{H}_I \big)_{qs}^{pr} & \text{ if } s \ne q\\
                                          -F_I \delta_{pr} & \text{ otherwise}
                                        \end{array}\right.\,.\label{eq:H_tilde}
\end{eqnarray}
The notation $A \big|_I$ indicates the $n \times n$ submatrix of $A$ whose rows are in the multi-index $I$,
$A \overset{q}{\leftarrow}b$ represents the matrix $A$ with the $q$-th column replaced by $b$,
and $e_p$ is the $p$-th element of the canonical basis of $\real^M$.
The condition $\eta_{\lozenge U} \in H_U$ can be imposed by extending the linear system with the $n^2$ equations $U^T\eta_{\lozenge U} = 0$, and solving it with a least-square subroutine.

To see how the quantities $\matG_I$ and $\mathbf{H}_I$ appear, note that
\begin{eqnarray}\label{eq:derdet_Jacobi}
  \frac{\partial F_I}{\partial U_q^p}(U) &=& 
  \frac{\partial \text{det}(U\big|_I)}{\partial U_q^p}(U)\nonumber\\
  &=& \text{tr}\left( \text{adj}(U\big|_I) \frac{\partial(U|_I)}{\partial U_q^p} (U\big) \right)\nonumber\\
  &=& \text{tr}
  \begin{pmatrix}
    \text{det}(U \overset{1}{\leftarrow} \delta_{1q}e_p)\big|_I & \dots & \text{det}(U \overset{1}{\leftarrow} \delta_{nq}e_p)\big|_I\\
    \vdots & \ddots & \vdots\\
    \text{det}(U \overset{n}{\leftarrow} \delta_{1q}e_p)\big|_I & \dots & \text{det}(U \overset{n}{\leftarrow} \delta_{nq}e_p)\big|_I
  \end{pmatrix}\nonumber\\
  &=& \big(\matG_I\big)_q^p\,,
\end{eqnarray}
where $\text{adj}(W)$ represents the classical adjoint, or adjugate, of the matrix $W$.
The quantities $\mathbf{H}_I$ appear similarly from the directional derivative of the gradient in the left-hand side of Eq.~\eqref{eq:Absil_main_eq}.
Thus, the matrices $\matG_I$ and $\mathbf{H}_I$ are associated to the first and second derivatives of $f_\lozenge$.
The factor $-F_I$ in the diagonal entries of $\mathbf{\tilde{H}}_I$
is a contribution from the projector $\Pi_{\cdot{}_\perp}$ to the directional derivative
(see Eq.~\eqref{eq:Absil_main_eq}).


One has to calculate several $n \times n$ determinants in this algorithm,
as it is necessary for the basis transformation of the wave function $\ket{\extWF}$,
see Eq.~\eqref{eq:FCI_transf}.
However, in the case of the present algorithm the number of such determinants is smaller
(see Sec.~\ref{sec:comp_cost})
and the external wave function is expressed always in the same basis,
with the CI coefficients that appear in the equations being the same at all iterations.

\subsubsection{\label{sec:alg_CISD}The case of a CISD wave function and the usage of symmetry adapted spatial orbitals}


We apply the equations outlined above for the case where the external wave function is a CISD
(configuration interaction with single and double excitations) wave function,
$\ket{\CI}$.
This exemplifies the advantages of this algorithm:
first, the summations over $I$ in Eq.~\eqref{eq:gen_lin_system_Absil} are always over the same single and double excitations,
with the same CI coefficients throughout the iterations.
This does not happen for the algorithm based on orbital rotations,
described in Sec.~\ref{sec:alg_orb_rot}.
In that case,
the wave function $\ket{\CI}$ must be given as a linear combination of excitations on top of the Slater determinant of each iteration,
that obviously changes,
and it is not necessarily the reference used to construct the CISD wave function.


A second advantage is that an efficient implementation is possible,
that benefits from the structure of the wave function,
in particular by exploiting symmetry adapted spatial orbitals.
In such case the matrix $U$ is divided into blocks
and so are the matrices used to calculate $F_I$, $\matG_I$, and $\mathbf{H}_I$.
Because the determinant of a block diagonal matrix is the product of the determinant of each block,
these quantities are decomposed in similar quantities for each spin and irreducible representation.
For example, if all indices $pqrs$ belong to the same spin and irreducible representation,
say alpha orbitals and irreducible representation $\irp=1$, then:
\begin{equation}
  \big( \mathbf{H}_I \big)^{pr}_{qs} = \big( \mathbf{H}_{I_1^\alpha} \big)^{pr}_{qs}
  \prod_{\irp \ne 1} F_{I_\irp^\alpha}
  \prod_{\irp} F_{I_\irp^\beta}\,,
\end{equation}
whereas if indices $pq$ belong to alpha orbitals, but $rs$ belong to beta orbitals (and same irreducible representation $\irp=1$):
\begin{equation}
  \big( \mathbf{H}_I \big)^{pr}_{qs} = \big( \matG_{I_1^\alpha} \big)^{p}_{q}
  \big( \matG_{I_1^\beta} \big)^{r}_{s}
  \prod_{\irp \ne 1} F_{I_\irp^\alpha} F_{I_\irp^\beta}\,.
\end{equation}
The complete equations for spin-restricted CISD wave functions are given and discussed in the Appendix.

\subsection{\label{sec:theor_comp}Theoretical comparison of the algorithms}

\subsubsection{\label{sec:alg_are_equivalent}The procedures are equivalent}


For every initial Slater determinant $\ket{\Phi_0}$, both algorithms produce the same Newton step,
$\ket{\Phi_0} \to \ket{\Phi}$.
In fact, first note that Algorithm 2, based on the coefficients matrices $U$,
takes into account the intrinsic geometry of the Grassmannian as studied by Absil and coworkers \cite{absilAAM04_80_199}.
Thus, it is independent of the chosen orbital basis, either for the complete one-particle space $\orbSp$ or for the vector space $[\ket{\Phi_0}]$.
Therefore once an orbital basis for $\ket{\Phi_0}$ has been chosen,
\begin{equation}
  \ket{\Phi_0} = \phi_1 \w \phi_2 \w \dots \w \phi_n\,,
\end{equation}
one can always consider a basis for $\orbSp$ that contains it,
and extend it to the virtual space of $\ket{\Phi_0}$:
\begin{equation}
  \orbSp = [\phi_1 \w \phi_2 \w \dots \w \phi_n \w \dots \w \phi_M]\,.
\end{equation}
In this condition, the matrix that represents $\ket{\Phi_0}$ is trivial:
\begin{equation}\label{eq:specific_U_refBasis}
  U =
  \begin{pmatrix}
    \mathbf{1}_{n \times n}\\
    \mathbf{0}_{(M-n) \times n}
  \end{pmatrix}\,.
\end{equation}

Now, let $\ket{\extWF}$ be expanded in the basis $\{\phi_p\}_{p=1}^M$,
such that both algorithms can be applied.
This is not required for Algorithm 2 but we consider it here for the sake of the present argument.
A straightforward application of Eq.~\eqref{eq:gen_lin_system_Absil} with $U$ given by Eq.~\eqref{eq:specific_U_refBasis} shows that the Jacobian and the Hessian in the Algorithm 1,
based on orbital rotations,
appear as submatrices at the right- and left-hand side matrices:
\begin{equation}
  \Jabsil = 
  \begin{pmatrix}
    \mathbf{0}_{n \times n}\\
    \Jac
  \end{pmatrix}
\end{equation}
\begin{equation}
  \Habsil^p_q =
  \begin{pmatrix}
    \mathbf{a}_{n \times n}\\
    \Hess_q^p
  \end{pmatrix}\,.
\end{equation}
Submatrix $\mathbf{a}$, although nonzero,
does not affect the solution of equation \eqref{eq:gen_lin_system_Absil},
since all columns of $\eta_{\lozenge U} \in H_U$ must be orthogonal to the columns of $U$ ($\eta_{\lozenge U} \in H_U$, see comment before Eq.~\eqref{eq:Absil_main_eq}):
\begin{equation}\label{eq:eta_proof_equivalence}
  \eta_{\lozenge U} =
  \begin{pmatrix}
    \mathbf{0}_{n \times n}\\
    \Kmat
  \end{pmatrix} = \mathcal{U} \Sigma \mathcal{V}^T\,,
\end{equation}
where we recall that a singular value decomposition will be applied to $\eta_{\lozenge U}$ in Algorithm 2.
Submatrix $\Kmat$ in Eq.~\eqref{eq:eta_proof_equivalence} is the same that solves Eq.~\eqref{eq:orb_rot_Newton_step} of Algorithm 1.
Finally, by expanding the functions exponential, cosine, and sine for matrices,
one shows that the updated $U$ matrix obtained by equation \eqref{eq:Absil_SVD} is the same as the first $n$ columns of Eq.~\eqref{eq:def_new_basis_orb_rot},
except by multiplication to the orthogonal matrix $\mathcal{V}^T$ (from the singular value decomposition),
that does not change the corresponding Slater determinant:
\begin{equation}
  [\ket{\Phi}] = [e^{\hat{K}}\ket{\Phi_0}] = \text{span}(U \mathcal{V} \cos \Sigma + \mathcal{U} \sin \Sigma)\,.
\end{equation}
Hence, both procedures are equivalent.

\subsubsection{\label{sec:comp_cost}Computational cost of the algorithms}

The algorithms presented above scale differently with the system size.
They both rely on the calculation of a large number of $n \times n$ determinants,
with $n$ being the number of electrons (compare Eq.~\eqref{eq:trans_fci_orbital_basis} with Eq.~\eqref{eq:def_FI}-\eqref{eq:def_HI}).
However, the number of such determinants is quite different in each case,
and they are compared in Table~\ref{tab:com_numb_det}.
Algorithm 1 scales exponentially, irrespective of the kind of external wave function,
since a full orbital transformation has to be performed.
On the other hand, in Algorithm 2, the basis used to expand the external wave function is preserved along the iterations,
and thus the number of Slater determinants used for its expansion remains unchanged ($N$).
This might be much smaller than the total number of possible Slater determinants ($N_\text{full}$),
as for a CISD external wave function.
However, even for a full configuration interaction (FCI) external wave function the number of determinant calculations that need to be performed increases slower than in Algorithm 1, although still exponentially.

\renewcommand{\arraystretch}{1.5}
\begin{table}
  \caption{\label{tab:com_numb_det}Number of distinct $n \times n$ determinants used in both algorithms,
    without considering any spin or spatial symmetry restriction.
    The number of electrons is represented by $n$,
    and the total number of spin-orbitals by $K$.
    The total number of possible Slater determinants is $N_\text{full}$,
    whereas $N$ represents the number of Slater determinants with non-vanishing contribution to the wave function in question.
  $N_\text{full}$ is assumed to grow as $\frac{1}{n}(\frac{K}{n})^K$ \cite{helgaker00_molec}.}
  \begin{tabular}{llllll}
    \hline
    Alg. 1 &  & $N_\text{full}^2 = {K\choose n}^2 = \big(\frac{K!}{n!(K-n)!}\big)^2 \sim \frac{1}{n^2}(\frac{K}{n})^{2K}$ \\
    Alg. 2 & general$\quad$ & $N(1 + nK + (nK)^2) \sim Nn^2K^2$ \\
           & FCI & $\big(\frac{K!}{n!(K-n)!}\big)(1 + nK + (nK)^2) \sim (\frac{K}{n})^KnK^2$\\
           &  \multirow{2}{*}{CISD} &$\big(1 + n(K-n) + \frac{n(n-1)(K-n)(K-n-1)}{4}\big)$\\
           & & $\quad\times(1 + nK + (nK)^2) \sim n^4K^4$\\
    \hline
  \end{tabular}
\end{table}

In actual computations, the external wave function is usually eigenfunction of $S_z$,
and often considers molecular spatial symmetry,
reducing the number of determinant calculations.
Relations among $F_I$, $\matG_I$, and $\mathbf{H}_I$ of similar multi-indices can also be used to avoid calculating several of the determinants, speeding up computations (see the Appendix~\ref{sec:app_CISD}).
Furthermore, the transformation of the wave function in Algorithm 1 can be performed much efficiently by the procedure proposed by Malmqvist \cite{malmqvistIJQC86_XXX_479}.
However, this analysis shows that a faster procedure can be obtained with Algorithm 2,
in particular because it can exploit the structure of the external wave function.

\subsubsection{\label{sec:comp_Zhang_Kollar}Comparison with the algorithm of Zhang and Kollar}

A possible pitfall of the Newton method as applied here is that it might converge to a saddle point of the overlap function,
as it strongly depends on the initial Slater determinant
(this is further explored in Sec.~\ref{sec:opt_ex_H2}).
In our test applications, discussed in Sec.~\ref{sec:examples},
this has not been a problem,
since we are concerned with ground state wave functions,
for which the restricted Hartree-Fock wave function is a perfectly fine starting guess.
Furthermore, for these cases the present algorithm converges quite fast,
typically in three iterations.
On the other hand,
the algorithm proposed by Zhang and Kollar \cite{zhangPRA14_89_012504} is more robust,
optimizing the overlap function by working with one orbital at a time:
in each iteration,
all orbitals of the Slater determinant are fixed except for, say, $\phi_i$;
this orbital is updated such that the overlap function is maximized under the restriction that $\phi_i$ is orthogonal to all other orbitals of the Slater determinant.
In the next iteration the procedure is repeated with $\phi_{i+1}$ (now with $\phi_i$ fixed),
or back to $\phi_1$, cyclically, until convergence is obtained.
This procedure is guaranteed to converge to a maximum, although not necessarily to a global maximum \cite{zhangPRA14_89_012504}.
However, it takes a large number of iterations to converge,
especially after reaching the plateau where just small updates are made after each iteration (see Fig.~1 of \cite{zhangPRA14_89_012504}).
Thus, the present algorithms, based on the Newton method, and the one from Zhang and Kollar,
are complementary and could be used in conjunction:
the more robust, but slower, algorithm of Zhang and Kollar can be used to reach the region of the Grassmannian close to the optimum point
(where iterations lead to small variations of the overlap function),
and thereafter the Newton method, as presented here, used for a fast convergence towards the critical point.

\section{\label{sec:examples}Some examples}


We have coded pilot implementations of the algorithms discussed in the previous section by using Python.
Tensor contractions to generate the elements of Eq.~\eqref{eq:gen_lin_system_Absil} are straightforwardly implemented with NumPy \cite{oliphant-06_0_0,waltCSE11_13_22}.
A hand-crafted C/Fortran code can obviously speed up computations,
but the present implementations suffices for our initial purposes.
For Algorithm 2, based on coefficients matrices, two implementations have been coded:
one for a general $\ket{\extWF}$,
and one specific for $\ket{\extWF}$ being a CISD wave function.
The accuracy of the implementations has been checked by the following means:
\begin{enumerate}
\item The Jacobian and Hessian in the Algorithm 1
  have also been calculated by finite differences
  and compared to the analytical versions
  (Eq.~\eqref{eq:jac_alg_orb_rot} and~\eqref{eq:hess_alg_orb_rot});
\item We performed sanity checks on the Algorithm 2 to assure that the solution of Eq.~\eqref{eq:gen_lin_system_Absil}, $\eta_{\lozenge U}$,
  really satisfies the original Eq.~\eqref{eq:Absil_main_eq}, of Absil and coworkers,
  with derivatives calculated by finite differences;
\item Both implementations of the Algorithm 2 lead to the same matrices of Eq.~\eqref{eq:gen_lin_system_Absil};
\item All implementations give the same iterations when the same $\ket{\CI}$ is used,
  as required by the conclusion of Sec.~\ref{sec:alg_are_equivalent}.
\end{enumerate}
In this section we will describe some example calculations carried out with these implementations.

\subsection{\label{sec:opt_ex_H2}The hydrogen molecule in a minimal basis}

We start by searching the Slater determinant with largest overlap to the exact ground state wave function (the FCI wave function, $\ket{\Psi_{0}}$) for the hydrogen molecule described by a minimal basis set, as discussed in Sec.~\ref{sec:ex_hydrogen}.
The STO-3G basis set representation \cite{hehreJCP69_51_2657} have been used.
Analytic expressions to measure entanglement/correlation can be derived for this case \cite{schliemannPRA01_64_022303}.
It is still of profound physical importance
since the distance between $\ket{\Psi_0}$ and the Grassmannian reaches its maximum in the dissociation limit, with non-interacting electrons,
as recently discussed by Ding and Schilling \cite{dingJCTC20_16_4159}.
Here we will focus on the behavior of the overlap function.

\begin{figure*}
  \includegraphics[width=1.0\textwidth]{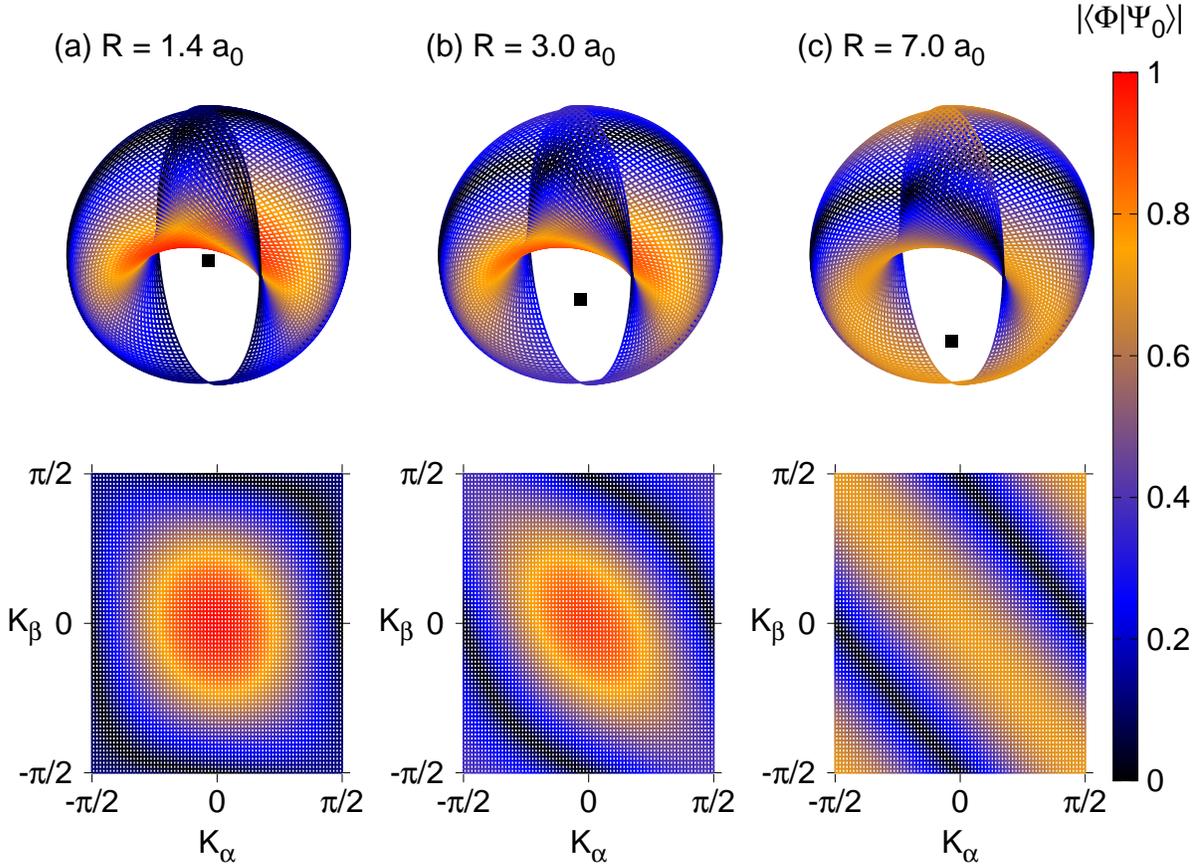}
  \caption{\label{fig:dist_to_fci_H2}The function $f_\square(K_\alpha, K_\beta) = |\bracket{\Phi(K_\alpha, K_\beta)}{\Psi_{0}}|$,
    for the hydrogen molecule, for the internuclear distances of
    (a) 1.4 a$_0$, (b) 3.0 a$_0$, and (c) 7.0 a$_0$.
    The upper images show the three-dimensional representation as discussed in Sec.~\ref{sec:ex_hydrogen},
    and the bottom images show the space of the parameters $K_\alpha$ and $K_\beta$.
    The exact wave function, $\ket{\Psi_{0}}$, is represented by a square in the upper images.}
\end{figure*}

For this small case there is a simple expression for the function to be optimized, being clear where its maximum is:
The exact wave function has the form of Eq.~\eqref{eq:doubly_exc_lin_comb_H2} and,
from Eq.~\eqref{eq:general_Phi_H2}, one obtains (considering normalized wave functions):
\begin{eqnarray}
  \bracket{\Phi(K_\alpha,K_\beta)}{\Psi_0} =&& C_0 \cos(K_\alpha) \cos(K_\beta)\nonumber\\
                                               &&+ \sqrt{1 - C_0^2} \sin(K_\alpha) \sin(K_\beta),
\end{eqnarray}
where $C_0$ is the coefficient of $\phi_+ \w \overline{\phi_+}$ (the ``reference determinant''),
and $\sqrt{1 - C_0^2}$ is the coefficient of $\phi_- \w \overline{\phi_-}$ (the ``doubly excited determinant'').
The absolute value of this overlap assumes its maximum at $\phi_+ \w \overline{\phi_+}$ ($K_\alpha = K_\beta = 0$),
if $|C_0| > \frac{1}{\sqrt{2}}$.
This is the case for every internuclear distance $R$,
in particular close to the equilibrium distance,
where the weight of the reference determinant is much larger than of the excited determinant.
See Fig.~\ref{fig:dist_to_fci_H2}.
However, note that $\phi_- \w \overline{\phi_-}$ ($|K_\sigma| = \pi/2$) is another critical point of this function,
and the Newton method might converge to it,
depending on the Slater determinant used to start the optimization.
In general, the overlap function might have several critical points over the Grassmannian,
and the optimization procedure might not converge to a maximum.
In practice, if $\ket{\Psi_0}$ is the exact or an approximate wave function for the ground state,
an obvious starting point is the Hartree-Fock Slater determinant.

The behavior of the overlap function depends on the distance between $\ket{\Psi_0}$ and the Grassmannian.
If $\ket{\Psi_0}$ is close to the Grassmannian, the maximum at $(K_\alpha,K_\beta) = (0,0)$ is very clear.
When $\ket{\Psi_0}$ becomes far from the Grassmannian,
moving away from  $\phi_+ \w \overline{\phi_+}$,
it gets closer to opposite regions of the Grassmannian,
in particular to $\phi_- \w \overline{\phi_-}$, whose contribution to $\ket{\Psi_0}$ increases.
In the limit $R \to \infty$, $|C_0| = \frac{1}{\sqrt{2}}$ and $\ket{\Psi_0}$ becomes equally distant to every Slater determinant of the form
\begin{eqnarray}
  \ket{\Phi} &=&
                 \left( \cos (K) \, \phi_+
                 + \sin (K) \, \phi_- \right)\nonumber\\
             &&\w \left( \cos (-K) \, \overline{\phi_+}
                 + \sin (-K) \, \overline{\phi_-} \right)\nonumber\\
             &=& \left( a \, \phi_+
                 + \sqrt{1-a^2} \, \phi_- \right)\nonumber\\
             &&\w \left( a \, \overline{\phi_+}
                 - \sqrt{1-a^2} \, \overline{\phi_-} \right)\,.
\end{eqnarray}
The maximum overlap is reached not at a single point of the Grassmannian,
but in a submanifold of it.
This submanifold,
seen in Fig.~\ref{fig:dist_to_fci_H2}.c),
is the stripe crossing diagonally the parameters space.
(Note that, although $\ket{\Psi_0}$ appears to be closer to $\phi_- \w \overline{\phi_-}$ in Fig.~\ref{fig:dist_to_fci_H2}.c,
this is just because this representation does not preserve the metric of the projective space.)

\subsection{\label{sec:opt_other_cases}Selected systems}

For H$_2$ in a minimal basis set ($1s$ orbital centered in each atom),
the point in the Grassmannian with maximum overlap with the exact ground state wave function is just the restricted Hartree-Fock Slater determinant,
$\ket{\HF} = \phi_+ \w \overline{\phi_+}$ (see Sec.~\ref{sec:opt_ex_H2}).
This is not the case in general.
In this subsection we present the value of $|\bracket{\minD}{\extWF}|$ for selected systems,
where $\ket{\minD}$ is the optimized Slater determinant with largest overlap with $\ket{\extWF}$ (and thus with minimum distance to $\ket{\extWF}$ according to the metrics in Eq.~\eqref{eq:metric_FubiniStudy}-\eqref{eq:metric_simple}).
As for $\ket{\extWF}$,
we use the configuration interaction with single and double excitations (CISD) wave function,
calculated with the Molpro package \cite{Molpro19}.
We consider the following basis sets:
STO-3G \cite{hehreJCP69_51_2657}, 6-31G \cite{hehreJCP72_56_2257},
and cc-pV$n$Z \cite{dunningjrJCP89_90_1007} ($n \in \{\text{D},\text{T},\text{Q}\}$).
The optimization procedure is started with $\ket{\HF}$,
the restricted Hartree-Fock wave function.
The frozen core approximation has been used in all examples;
This implies that core orbitals of $\ket{\minD}$ are the same of $\ket{\HF}$,
since the elements of the Jacobian associated to these orbitals are zero
(see Eq.~\eqref{eq:jac_alg_orb_rot}).
Thus, the optimization of $\ket{\minD}$ can be made over a Grassmannian of smaller order,
by considering only orbitals correlated in $\ket{\CI}$.

Besides applying the present algorithms,
the objective of this subsection is to compare $\ket{\minD}$ and $\ket{\HF}$,
or equivalently $|\bracket{\minD}{\CI}|$ and $|\bracket{\HF}{\CI}|$,
the latter being directly available after a CISD calculation.
Note that
\begin{eqnarray}\label{eq:approx_minD_HF}
  \bracket{\HF}{\CI} &=& \average{\HF}{\hat{P}_{\text{minD}} + \hat{Q}_{\text{minD}}}{\CI}\nonumber\\
   &=& \bracket{\HF}{\minD}\bracket{\minD}{\CI}\nonumber\\
           &&+\average{\HF}{\hat{Q}_{\text{minD}}}{\CI}\nonumber\\
  &\approx& \bracket{\HF}{\minD}\bracket{\minD}{\CI}\,,
\end{eqnarray}
where $\hat{P}_{\text{minD}} = \ketbra{\minD}{\minD}$ is the projector onto $\ket{\minD}$ and $\hat{Q}_{\text{minD}} = 1 - \hat{P}_{\text{minD}}$.
The approximation in Eq.~\eqref{eq:approx_minD_HF} holds whenever $\ket{\minD}$ is not too far from $\ket{\HF}$,
hence the term with $Q_{\text{minD}}$ can be neglected.
Thus:
\begin{equation}\label{eq:relation_ration_ovlpHFminD}
  \frac{\bracket{\HF}{\CI}}{\bracket{\minD}{\CI}}
  \approx \bracket{\minD}{\HF}\,.
\end{equation}
Recall that $|\bracket{\minD}{\HF}|$ is related to the distance
(measured in $\projective\extProd{}$) between these two Slater determinants,
and either side of Eq.~\eqref{eq:relation_ration_ovlpHFminD} can be used to estimate the importance
of using the optimized $\ket{\minD}$ (instead of plain $\ket{\HF}$) to measure the correlation.
In the calculations we carried out, Eq.~\eqref{eq:relation_ration_ovlpHFminD} holds.

\begin{figure*}
  \includegraphics[width=1.0\textwidth]{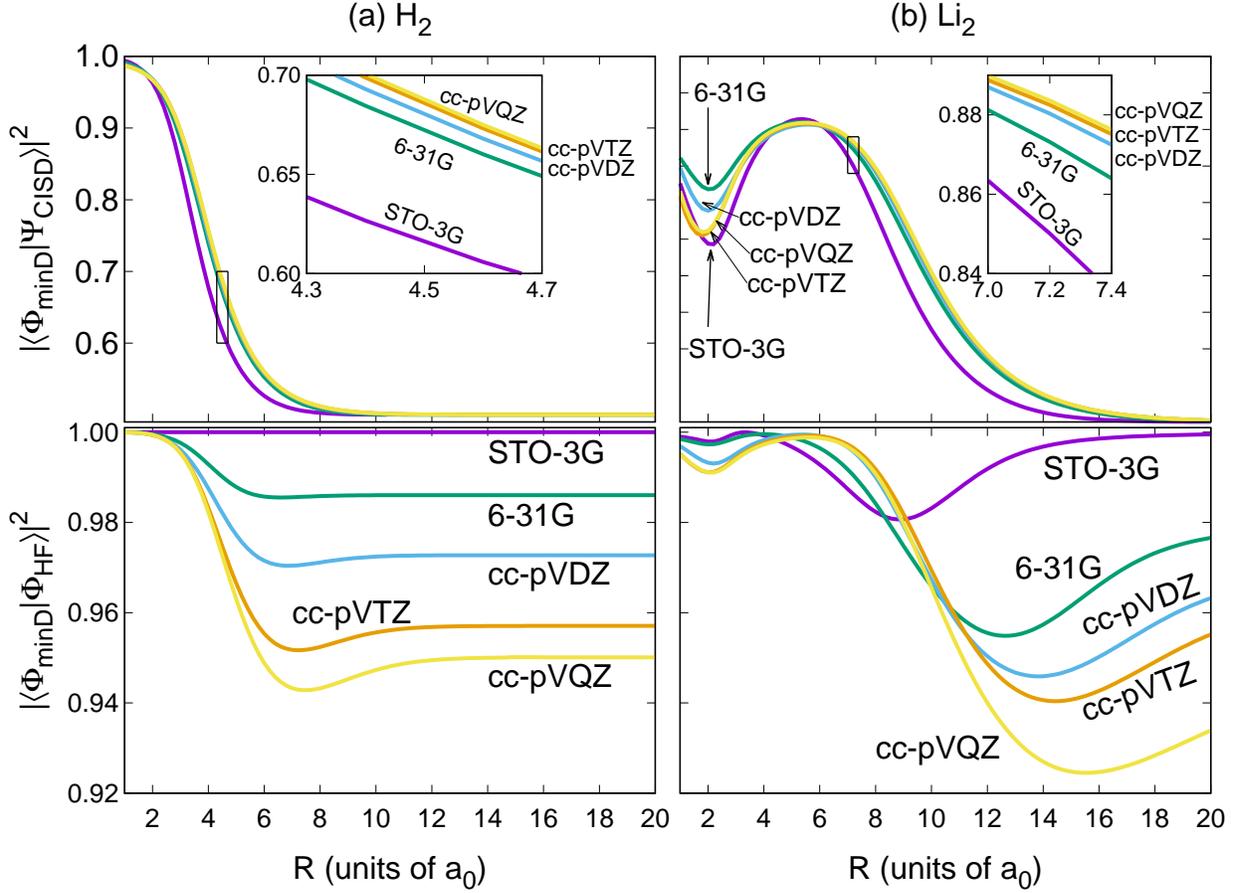}
  \caption{\label{fig:dGr_H2_Li2}The square of the absolute value for the overlap between $\ket{\minD}$, and $\ket{\CI}$ (top graphs) or $\ket{\HF}$ (bottom graphs).
  Inset graphs show the curves in a closer range at the region indicated by the rectangle.}
\end{figure*}


The first application is for H$_2$ in larger basis sets,
being $\ket{\CI}$ the exact wave function.
In Fig.~\ref{fig:dGr_H2_Li2}.a) we show the values of $|\bracket{\minD}{\CI}|^2$ and $|\bracket{\minD}{\HF}|^2$, as function of the internuclear distance.
The qualitative behavior of $|\bracket{\minD}{\Psi_0}|^2$ is the same at all the basis sets,
namely, it rapidly decreases during the dissociation, reaching the value of circa 0.5.
The reasoning has been discussed in Sec.~\ref{sec:opt_ex_H2}.
However, for larger basis sets $\ket{\minD}$ deviates from $\ket{\HF}$,
in particular at larger interatomic distances.
For the cc-pVQZ basis set at $R \approx 7.0\, a_0$, $|\bracket{\minD}{\Psi_0}|^2$ is 94\% of the value of
$|\bracket{\HF}{\Psi_0}|^2$.


For the Li$_2$ molecule, we observe a more complex variation of $|\bracket{\minD}{\Psi_0}|^2$
along the dissociation, presenting a maximum at $R \approx 5.5\,a_0$,
slightly after the equilibrium distance (at $R = 5\,a_0$).
This maximum indicates that the correlation at this point is minimum,
as already discussed by Benavides-Riveros and coworkers \cite{benavides-riverosPCCP17_19_12655}.
Although the qualitative behavior of $|\bracket{\minD}{\Psi_0}|^2$ does not change with basis set,
$|\bracket{\minD}{\HF}|^2$ is strongly dependent on the basis set.
The deviation of $|\bracket{\minD}{\Psi_0}|^2$ from $|\bracket{\HF}{\Psi_0}|^2$ increases for larger basis sets, particularly in regions where correlation is also large.
Note that, under the frozen core approximation, the CISD wave function is also exact for Li$_2$.

Table \ref{tab:large_sys} shows results from the optimization of $\ket{\minD}$, with respect to the CISD wave function, for some other molecules.
As it happens for the previous examples,
$\ket{\minD}$ is quite close to $\ket{\HF}$, especially for systems with small correlation,
such as water in the equilibrium geometry.
For the stretched water molecule correlation effects increase
and $|\bracket{\minD}{\HF}|^2$ decreases, but it is still over 0.98.
Observe that $\bracket{\minD}{\HF}$ is rather less sensitive to the increase of basis set than
$\bracket{\minD}{\CI}$.
For the ozone molecule, and for the three transition metal diatomic molecules,
both $\bracket{\minD}{\CI}$ and $\bracket{\minD}{\HF}$ are quite insensitive with basis set.

\renewcommand{\arraystretch}{1.5}
\begin{table*}
  \caption{Results for the optimization of the Slater determinant with the largest overlap with
  the CISD wave function ($\ket{\minD}$), for selected systems and basis sets.}
  \label{tab:large_sys}
  \begin{tabular}{lllcc}
    \hline
    molecule & geometry & basis set
    & $|\bracket{\minD}{\CI}|^2 \times 100$
    & $|\bracket{\minD}{\HF}|^2 \times 100$\\
    \hline
    H$_2$O & $R_{\text{OH}} = 0.9633$ \AA
      & cc-pVDZ            & 95.063 & 99.961\\
    & $a_{\text{HOH}} = 102.57^\circ$
       & cc-pVTZ            & 94.504 & 99.954\\
    &  & cc-pVQZ            & 94.391 & 99.945\\
    \hline
    H$_2$O & $R_{\text{OH}} = 2.5$ \AA
       & cc-pVDZ            & 63.356 & 98.533\\
    & $a_{\text{HOH}} = 102.57^\circ$
       & cc-pVTZ            & 70.812 & 98.481\\
    &  & cc-pVQZ            & 72.786 & 98.518\\
    \hline
    O$_3$ & $R_{\text{OO}} = 1.2728$ \AA
       & cc-pVDZ            & 87.310 & 99.405\\
    & $a_{\text{OOO}} = 116.75^\circ$
       & cc-pVTZ            & 87.181 & 99.539\\
    &  & cc-pVQZ            & 87.215 & 99.572\\
    \hline
    ScH & $R_{\text{ScH}} = 1.7754$ $a_0$
       & cc-pVDZ            & 92.059 & 99.785\\
    &  & cc-pVTZ            & 92.361 & 99.769\\
    &  & cc-pVQZ            & 92.472 & 99.769\\
    \hline
    CuH & $R_{\text{CuH}} = 1.4626$ $a_0$
       & cc-pVDZ            & 93.451 & 99.722\\
    &  & cc-pVTZ            & 93.544 & 99.761\\
    &  & cc-pVQZ            & 93.481 & 99.761\\
    \hline
    ZnO & $R_{\text{ZnO}} = 1.7047$ $a_0$
       & cc-pVDZ & 92.016 & 99.593\\
    &  & cc-pVTZ & 91.916 & 99.698\\
    &  & cc-pVQZ & 91.827 & 99.723\\
    \hline
  \end{tabular}
\end{table*}

\section{\label{sec:conclusions}Conclusions}


In this article we described procedures to optimize the critical points of the overlap to an arbitrary wave function over the set of Slater determinants.
This can be used to measure the distance between a correlated wave function and the set of Slater determinants,
that is a measure of correlation incorporated in the wave function.
Obtaining such distance is important to understand the relation between electronic correlation and
entanglement \cite{benavides-riverosPRA17_95_032507,dingJCTC20_16_4159},
and to analyze the interplay between static and dynamic correlation
\cite{benavides-riverosPCCP17_19_12655}.
The optimization procedures described here can be used for relatively large systems,
using exact as well as approximate wave functions.
An efficient version specific for a configuration interaction with single and double excitations (CISD) wave function is presented.
We developed these procedures by acknowledging that the set of Slater determinants form a submanifold
of the space of wave functions.
This manifold is the \emph{Grassmannian},
whose geometry is of central importance in mathematics,
but still of few known applications in atomic and molecular physics or in theoretical chemistry,
even though Slater determinants are key elements to electronic structure theory.
This work shows how the geometry of the Grassmannian can be used
for both theoretical considerations on the electron correlation and the practical optimization of a Slater determinant.


We have considered two approaches for an optimization process based on the Newton method.
The first is using the standard parametrization by orbital rotations,
and the second is an algorithm that explores the intrinsic geometry of the Grassmannian,
as described by Absil et al \cite{absilAAM04_80_199}.
We showed that both algorithms are equivalent, in the sense that they lead to the same
iterations (the same sequence of Slater determinants).
However, the second algorithm allows a much more efficient implementation,
since it avoids the undesirable basis transformation step of the external wave function,
that is not only time consuming,
but destroys the original rank truncation of the wave function.
Furthermore, the second algorithm uses directly the full coefficients matrix of the orbitals
in the Slater determinant,
that is a set of non-independent parameters.
It is often assumed that wave function optimizations based on the Newton method cannot be performed in such condition,
and an independent set of parameters that covers the desired space of wave functions is necessary.
Here we showed that this is perfectly possible,
as long as this is made carefully to project out the variations on the redundant parameters that do not change the wave function.
This is done after considerations on the geometry of the underlying manifold \cite{absilAAM04_80_199}.

The present algorithms converge quickly, typically in 3 iterations,
as long that the starting point is reasonably close to the maximum overlap Slater determinant,
such as the restricted Hartree-Fock Slater determinant for most of the systems.
However, the procedures might converge to a relative maximum or to saddle points, if started with a poor initial guess.
For such difficult cases,
the present algorithms can be used in conjunction with the procedure of Zhang and Kollar
\cite{zhangPRA14_89_012504},
that converges more robustly, although in much more iterations.

Applications of the algorithm suggest that using the restricted Hartree-Fock wave function, $\ket{\HF}$, to measure correlation is qualitatively equivalent to using $\ket{\minD}$,
the Slater determinant that minimizes the distance to an external wave function $\ket{\extWF}$.
For most of the cases, $|\bracket{\minD}{\extWF}|^2$ accounts for more than 99\% of $|\bracket{\HF}{\extWF}|^2$.
However, there are quantitative differences when correlation is very large,
and basis set truncation effect might be strong on both $|\bracket{\minD}{\extWF}|$ and $|\bracket{\minD}{\HF}|$.
We emphasize that the present examples are all singlet and closed shell systems,
with CISD wave functions based on a restricted Hartree-Fock reference,
what forces $\ket{\minD}$ to be also spin restricted,
and thus naturally close to $\ket{\HF}$.
For instance, $\ket{\minD}$ is actually very far from the unrestricted Hartree-Fock wave function in cases of instabilities on the restricted Hartree-Fock wave function,
as in the dissociation limit of H$_2$ and Li$_2$ molecules discussed here.
Larger differences between $|\bracket{\minD}{\CI}|$ and $|\bracket{\HF}{\CI}|$ are thus expected for open shell cases,
where the spin restriction over $\ket{\minD}$ has to be relaxed.
Furthermore, the single-reference CISD method is of limited usage nowadays,
and the evaluation of $|\bracket{\minD}{\CC}|$,
where $\ket{\CC}$ is the coupled-cluster with single and double excitations
\cite{purvisJCP82_76_1910,bartlettRMP07_79_291},
for instance, is more appealing.
However, the full set of excited determinants would be needed for the present algorithms.
One possible approximation is to consider only the projection of the CCSD,
wave function into the space of up to doubly excited determinants, that is a CISD-like wave function.
Hence, the present algorithms allow several numerical investigations on the quantification of electronic correlation.

\appendix

\section{Appendices}

\subsection{Spatial orbitals and symmetry considerations}

In this appendix we present explicit formulas for the case where $\ket{\extWF}$ is constructed from symmetry adapted spatial orbitals, based on Abelian point groups.
This means that the spin-orbital space (Eq.~\eqref{eq:orb_space}) is given as the following direct sum of spaces:
\begin{equation}
  \label{eq:orb_space_symm}
  \orbSp = \orbSp_1^\alpha \oplus \dots \oplus \orbSp_g^\alpha
  \oplus \orbSp_1^\beta \oplus \dots \oplus \orbSp_g^\beta\,,
\end{equation}
where the vector space associated to the irreducible representation (irrep) $\irp$ and spin $\sigma$ is of dimension $M_\irp$:
\begin{equation}
  \orbSp_\irp^\sigma = [\phi_1^\irp\otimes\sigma \w \dots \w \phi_{M_\irp}^\irp\otimes\sigma]\,.
\end{equation}

In this condition, Eq.~\eqref{eq:def_orb_rot_parameters} becomes \cite{helgaker00_molec}:
\begin{equation}\label{eq:def_orb_rot_parameters_sym}
  \hat{K} = \sum_\irp \sum_{(i,a)\in\irp} K_i^{a,\irp} \left( E_i^{a,\irp} - E_a^{i,\irp} \right)\,,
\end{equation}
where
$E_q^{p,\irp} =  a_{\alpha p, \irp}^\dagger a_{\alpha q,\irp} + a_{\beta p,\irp}^\dagger a_{\beta q,\irp}$
are the singlet excitation operators for the irrep $\irp$.
Extension of the algorithm based in orbital rotations discussed in Sec.~\ref{sec:alg_orb_rot} is straightforward, although care should be taken to the orbital ordering and the sign of coefficients.

\subsubsection{\label{sec:app_gen_WF}Algorithm 2: equations for a general $\ket{\extWF}$}

Function $f_\lozenge$ becomes:
\begin{equation}
  \label{eq:gen_f}
  f_\lozenge(U) = 
  \sum_{\substack{I\\\text{occ}(I) = \text{occ}(U)}} C_I \,
  \prod_\irp F_{I_\alpha^\irp} F_{I_\beta^\irp}\,,
\end{equation}
and Eq.~\eqref{eq:gen_lin_system_Absil} becomes:
\begin{equation}
  \big( \Habsil_{\sigma \irp}^{\sigma' \irpP} \big)^{pr}_{qs} \big( \big(\eta_{\lozenge U}\big)_{\sigma' \irpP} \big)^r_s
  = -\big( \Jabsil_{\sigma \irp} \big)^p_q\,,
\end{equation}
where ${\sigma \irp}$ indicates the block of the corresponding matrix associated to spin $\sigma$ and irrep $\irp$.
The notation ``$\text{occ}(I) = \text{occ}(U)$'' indicates that only terms of $\ket{\extWF}$ that have the same number of electrons as $U$ in all blocks must be included.
Indices $p$ and $r$ run over all orbitals of that symmetry ($M_\irp$ in number),
whereas $q$ and $s$ run over the electrons in that spin and symmetry.
In the following equations, quantities $\Jabsil_{\sigma \irp}$ and $\matG_{I_\sigma^\irp}$
are two-index tensors of shape $(M_\irp, n_\irp^\sigma)$,
whereas $\Habsil_{\sigma \irp}^{\sigma' \irpP}$ and $\mathbf{\tilde{H}}_{I_\sigma^\irp}$ are four-index quantities, of shape $(M_\irp,n_\irp^\sigma,M_\irp,n_\irpP^{\sigma'})$ and $(M_\irp,n_\irp^\sigma,M_\irp,n_\irp^\sigma)$, respectively.
The tensor product $\otimes$ between a $(M, n)$ quantity by a $(M', n')$ quantity is the $(M,n,M',n')$ quantity whose entries are:
\begin{equation}
  \big( A \otimes B \big)^{pr}_{qs} = A^p_qB^r_s\,.
\end{equation}

\begin{equation}\label{eq:C_sym}
  \Jabsil_{\sigma \irp} =
  \Pi_{U_\sigma^\irp \perp}
  \sum_{\substack{I\\\text{occ}(I) = \text{occ}(U)}} C_I
  \left( \prod_{\{\sigma', \irpP\} \ne \{\sigma, \irp\}} F_{I_{\sigma'}^\irpP} \right)
  \matG_{I_\sigma^\irp}
\end{equation}

\begin{equation}\label{eq:X_sym}
  \Habsil_{\sigma \irp}^{\sigma \irp}=
  (\Pi_{U_\sigma^\irp \perp} \otimes \mathbb{1})
  \sum_{\substack{I\\\text{occ}(I) = \text{occ}(U)}} C_I
  \left( \prod_{\{\sigma', \irpP\} \ne \{\sigma, \irp\}} F_{I_{\sigma'}^\irpP} \right)
  \mathbf{\tilde{H}}_{I_\sigma^\irp}
\end{equation}

\begin{eqnarray}\label{eq:X_off_sym}
  \Habsil_{\sigma \irp}^{\sigma' \irpP} &=&
  \big(\Pi_{U_\sigma^\irp \perp} \otimes \Pi_{U_{\sigma'}^\irpP \perp} \big)\nonumber\\
  &&\sum_{\substack{I\\\text{occ}(I) = \text{occ}(U)}} C_I
  \left(
    \prod_{\substack{
        \{\sigma'', \irpPP\} \ne \{\sigma', \irpP\} \\
        \{\sigma'', \irpPP\} \ne \{\sigma , \irp \}}}
    F_{I_{\sigma''}^{\irpPP}} \right)
  \matG_{I_\sigma^\irp} \otimes \matG_{I_{\sigma'}^\irpP}\,,\nonumber\\
\end{eqnarray}
where the last equation holds for ${\{\sigma,\irp\} \ne \{\sigma',\irpP\}}$.
These equations are obtained after considering the block diagonal structure of the matrices $U\big|_I$,
$\left( U \overset{q}{\leftarrow} e_p \right) \big|_I$, and
$\left( U \overset{q}{\leftarrow} e_p \overset{s}{\leftarrow} e_r \right) \big|_I$,
along with the fact that the determinant of a block diagonal matrix is the product of the determinants of its blocks (see Sec.~\ref{sec:alg_CISD}).
If $I$ has a different number of electrons than in $U$ in any of its $\sigma\irp$ block,
(i.e, $\text{occ}(I) \ne \text{occ}(U)$)
the matrices above have non-square blocks and their determinants are zero.

\subsubsection{\label{sec:app_CISD}Algorithm 2: equations for a restricted CISD wave function}

Suppose now that $\ket{\extWF}$ is a spin restricted CISD wave function, based on a restricted and closed shell reference Slater determinant.
The reference determinant is given as:
\begin{eqnarray}
  \ketbig{\Phi_0}
  &=& \phi_1^{\irp=1} \w
      \dots \w \phi_{n_g}^{\irp=g} \w
      \overline{\phi}_1^{\irp=1} \w
      \dots \w \overline{\phi}_{n_g}^{\irp=g}\nonumber\\
  &=& \ketbig{\Phi_0}_1 \w
      \dots \w \ketbig{\Phi_0}_g \w
      \ketbig{\overline{\Phi}_0}_1 \w
      \dots \w \ketbig{\overline{\Phi}_0}_g\,,
\end{eqnarray}
where, for example, the subspace of $[\ket{\Phi_0}]$ associated to alpha orbitals of irrep $1$ is $[\ket{\Phi_0}_1]$ and so on.
Overlines indicate beta spin.
To simplify the notation, only the blocks where some excitation occurs will be shown,
and the blocks not shown are assumed to be equal as in the reference determinant.
Thus, for example:
\begin{eqnarray}
  \ketbig{\Phi_i^a}_\irp =&&
  \ketbig{\Phi_0}_1 \w \dots \w \phi_1^\irp \w \dots \w \phi_{i-1}^\irp\nonumber\\
  &&\w \phi_{i+1}^\irp \w \dots \w \phi_{n_\irp}^\irp \w \phi_a^\irp \w \dots \w \ketbig{\overline{\Phi}_0}_g
\end{eqnarray}
is an alpha single excitation from $i$ to $a$ in the irrep $\irp$.
With this notation, the CISD wave function can be written as:
\begin{equation}
  \label{eq:CISDwf}
  \begin{split}
    \ket{\CI} =
    & C_0 \ketbig{\Phi_0}\\
    & + \sum_\irp\sum_{(i,a) \in \irp} C_{i}^{a, \irp}
    \Big(
    \ketbig{\Phi_i^a}_\irp
    + \ketbig{\overline{\Phi}_i^a}_\irp
    \Big)\\
    & + \sum_\irp\sumijabrestr C_{ij}^{ab, \irp}
    \Big(
    \ketbig{\Phi_{ij}^{ab}}_\irp
    + \ketbig{\overline{\Phi}_{ij}^{ab}}_\irp
    \Big)\\
    & + \sum_\irp\sumijabfull\Dss_{ij}^{ab, \irp} \,
    \ketbig{\Phi_{i}^{a}}_\irp
    \dots
    \ketbig{\overline{\Phi}_{j}^{b}}_\irp\\
    & + \sum_{\irp > \irpP} \sumijabmix \Dmixaa_{ij}^{ab, \irp\irpP}
    \Big(
    \ketbig{\Phi_{j}^{b}}_\irpP
    \dots
    \ketbig{\Phi_{i}^{a}}_\irp\\
    &\quad\quad+ \ketbig{\overline{\Phi}_{j}^{b}}_\irpP
    \dots
    \ketbig{\overline{\Phi}_{i}^{a}}_\irp
    \Big)\\
    &  + \sum_{\irp > \irpP}\sumijabmix \Dmixab_{ij}^{ab, \irp\irpP}
    \Big(
    \ketbig{\Phi_{j}^{b}}_\irpP
    \dots
    \ketbig{\overline{\Phi}_{i}^{a}}_\irp\\
    &\quad\quad+ \ketbig{\Phi_{i}^{a}}_\irp
    \dots
    \ketbig{\overline{\Phi}_{j}^{b}}_\irpP
    \Big)\\
    & \sum_{\substack{I\text{ doubles over }\ket{\Phi_0}\\\text{occ}(I) \ne \text{occ}(0)}}
    C_I \ketbig{\Phi_I}\,,
  \end{split}
\end{equation}
where $\Dss_{ij}^{ab, \irp} =  \Dss_{ji}^{ba, \irp}$, as it is a restricted wave function.
As will be seen below,
$\Dmixaa_{ij}^{ab, \irp\irpP}$ and $\Dmixab_{ij}^{ab, \irp\irpP}$ always appear summed,
and thus the coefficients of double excitations arising as product of single excitations at different blocks are merged in a single quantity $\Dmix$:
\begin{eqnarray}
  \Dmix_{ij}^{ab, \irp\irp} &=& \Dss_{ij}^{ab, \irp}\\
  \Dmix_{ij}^{ab, \irp\irpP} &=& \Dmixaa_{ij}^{ab, \irp\irpP} + \Dmixab_{ij}^{ab, \irp\irpP}
                            \quad\text{ for }\irp \ne \irpP\,.
\end{eqnarray}
Determinants that have some spin/irrep with a number of electrons different than in the reference determinant are collected in the last term,
and they contribute neither to $f_\lozenge$ nor to the matrices used in the optimization.
Applying this wave function in the equations of Sec.~\ref{sec:app_gen_WF},
the following equations are obtained.
At first we define some intermediates:
\begin{eqnarray}
  \mathcal{F}_0 &=& \prod_\irp F_{I_0^\irp}^2\\
  \mathcal{F}_0^\irp &=& \prod_{\irpP \ne \irp} F_{I_0^\irpP}^2\\
  \mathcal{F}_0^{\irp\irpP} &=&
    \prod_{\substack{{\irpPP \ne \irp}\\{\irpPP \ne \irpP}}} F_{I_0^{\irpPP}}^2\,,
\end{eqnarray}
with analogous definitions for $\mathcal{F}_0^{\irp\irpP\irpPP}$ and $\mathcal{F}_0^{\irp\irpP\irpPP\irpPPP}$;
\begin{equation}
  \label{eq:K_for_CISD}
  \mathcal{K}^{\irp\irpP} = 2 \sumijabmix \Dmix_{ij}^{ab, \irp\irpP}
  F_{I_0^\irp} F_{I_0^\irpP} F_{I_i^{a,\irp}} F_{I_j^{b,\irpP}}
  \quad\text{ for }\irp \ne \irpP\,.
\end{equation}
In these equations,
${I_0^\irp}$ is the multi-index of the reference for irrep $\irp$,
whereas ${I_i^{a,\irp}}$ is the multi-index for the single excitation from $i$ to $a$, also in $\irp$.
The other type of multi-index that appears is ${I_{ij}^{ab,\irp}}$, for double excitations at irrep $\irp$;
\begin{eqnarray}
  \label{eq:L_for_CISD}
  \mathcal{L}^\irp =&& 2 F_{I_0^\irp}
  \left(
    \sum_{(i,a) \in \irp} C_{i}^{a, \irp} F_{I_i^{a,\irp}}
    + \sumijabrestr C_{ij}^{ab, \irp}  F_{I_{ij}^{ab,\irp}}
  \right)\nonumber\\
  &&+ \sumijabfull \Dmix_{ij}^{ab, \irp\irp}
  F_{I_i^{a,\irp}} F_{I_j^{b,\irp}}
\end{eqnarray}
The factor $2$ takes into account the contributions from alpha and beta excitations,
both totally within the same irrep $\irp$.
With the so far defined quantities, we are able to calculate $f_\lozenge(U)$, assuming $U$ orthonormal:
\begin{equation}
  \label{eq:f_for_CISD}
  f_\lozenge(U) = C_0 \mathcal{F}_0
  + \sum_\irp \mathcal{F}_0^\irp\mathcal{L}^\irp
  + \sum_{\irp > \irpP} \mathcal{F}_0^{\irp\irpP}\mathcal{K}^{\irp\irpP}\,.
\end{equation}
It is not difficult to see the origin of each term of Eq.~\eqref{eq:f_for_CISD}
(compare to Eq.~\eqref{eq:gen_f}):
The first is the contribution from the reference determinant, that is the product of all $F_{I_0^\irp}$, for all $\irp$ and for each spin.
Since it is a closed shell restricted wave function, this is just $\mathcal{F}_0$.
The second term of Eq.~\eqref{eq:f_for_CISD} is the contribution of all excitations within the same irrep:
for all such excitations, the other irrep blocks contribute with a $F_{I_0^\irpP}^2$, that form a common $\mathcal{F}_0^\irp$;
the contribution of the irrep in question is the CI coefficient, times an appropriate $F_{I^\irp}$, as can be seen in Eq.~\eqref{eq:L_for_CISD}.
The last term is the contribution of excitations at mixed irreps (say $\irp$ and $\irpP$):
each determinant contribute with one $F_{I_0^\irp}$, one $F_{I_0^\irpP}$ (from the spins where no excitations occurred, whichever they are), and the $F_{I^\irp}$ and $F_{I^\irpP}$ of corresponding single excitations.
This is clearly seen in \eqref{eq:K_for_CISD}, and the contribution of remaining irreps forms $\mathcal{F}_0^{\irp\irpP}$.

For the matrices $\Habsil$ and $\Jabsil$, Eq.~\eqref{eq:C_sym} to \eqref{eq:X_off_sym},
the interpretation is similar.
We define the intermediates
(the quantities $\matG_{I_i^{a,\irp}}$, $\hat{\matG}_{ia}^\irp$, and $\matM^\irp$ have shape $(M_\irp,n_\irp)$,
whereas $\bigG^\irp$, $\mathbf{\tilde{H}}$ and $\bigH^\irp$,
have shape $(M_\irp,n_\irp,M_\irp,n_\irp)$):
\begin{equation}
  \hat{\matG}_{ia}^\irp
  = F_{I_0^\irp} \matG_{I_i^{a,\irp}}
  + F_{I_i^{a,\irp}} \matG_{I_0^\irp}
\end{equation}

\begin{equation}\label{eq:def_Csingles_bar}
  \overline{C}_i^{a,\irp} = C_{i}^{a, \irp}
  + \sum_{\irpB \ne \irp} \frac{1}{F_{I_0^{\irpB}}}
  \sum_{(j,b) \in \irpB} F_{I_j^{b,\irpB}} \Dmix_{ij}^{ab, \irp\irpB}
\end{equation}

\begin{equation}
  \begin{split}
    \matM^\irp = &
    C_0 F_{I_0^\irp} \matG_{I_0^\irp}\\
    & + \sum_{(i,a) \in \irp} \overline{C}_i^{a,\irp}
    \hat{\matG}_{ia}^\irp \\
    & + \sumijabrestr C_{ij}^{ab, \irp}
    \Big(
        F_{I_0^\irp} \matG_{I_{ij}^{ab,\irp}} + F_{I_{ij}^{ab,\irp}} \matG_{I_0^\irp}
    \Big)\\
    & + \sumijabfull \Dmix_{ij}^{ab, \irp\irp}
    F_{I_i^{a,\irp}} \matG_{I_j^{b,\irp}}\\
  \end{split}
\end{equation}

\begin{equation}
  \begin{split}
    \bigG^\irp = &
    C_0 \matG_{I_0^\irp} \otimes \matG_{I_0^\irp}\\
    & + \sum_{(i,a) \in \irp} \overline{C}_i^{a,\irp}
    \Big(
        \matG_{I_i^{a,\irp}} \otimes \matG_{I_0^\irp}
        + \matG_{I_0^\irp} \otimes \matG_{I_i^{a,\irp}}
    \Big)\\
    & + \sumijabrestr C_{ij}^{ab, \irp}
    \Big(
        \matG_{I_0^\irp} \otimes \matG_{I_{ij}^{ab,\irp}}
        + \matG_{I_{ij}^{ab,\irp}} \otimes \matG_{I_0^\irp}
    \Big)\\
    & + \sumijabfull \Dmix_{ij}^{ab, \irp\irp}
    \matG_{I_i^{a,\irp}} \otimes \matG_{I_j^{b,\irp}}\\
  \end{split}
\end{equation}

\begin{equation}
  \begin{split}
    \bigH^\irp = &
    C_0 F_{I_0^\irp} \mathbf{\tilde{H}}_{I_0^\irp}\\
    & + \sum_{(i,a) \in \irp} \overline{C}_i^{a,\irp}
    \Big(
        F_{I_0^\irp} \mathbf{\tilde{H}}_{I_{i}^{a,\irp}}
        + F_{I_{i}^{a,\irp}} \mathbf{\tilde{H}}_{I_0^\irp}
    \Big)\\
    & + \sumijabrestr C_{ij}^{ab, \irp}
    \Big(
        F_{I_0^\irp} \mathbf{\tilde{H}}_{I_{ij}^{ab,\irp}}
        + F_{I_{ij}^{ab,\irp}} \mathbf{\tilde{H}}_{I_0^\irp}
    \Big)\\
    & + \sumijabfull \Dmix_{ij}^{ab, \irp\irp}
    F_{I_i^{a,\irp}} \mathbf{\tilde{H}}_{I_j^{b,\irp}}
  \end{split}
\end{equation}
Note that the terms from ``single excitations'' in these quantities contain a contribution from double excitations that occur partially in $\irp$ and partially in another irrep (see Eq.~\eqref{eq:def_Csingles_bar}),
and thus represent indeed single excitations from the point of view of irrep $\irp$.

Finally:
\begin{widetext}
\begin{equation}
  \Jabsil_\irp =
  \Pi_{U^\irp \perp} \Big\{
  \mathcal{F}_0^\irp \matM^\irp
  + \Big(
      \sumB \mathcal{F}_0^{\irp\irpB} \mathcal{L}^\irpB
      + \sumBB \mathcal{F}_0^{\irp\irpB\,\irpBB} \mathcal{K}^{\irpB\,\irpBB}
  \Big) F_{I_0^\irp} \matG_{I_0^\irp}
  \Big\}\\
\end{equation}
\begin{equation}
  \begin{split}
    \Habsil_\irp^\irp = &\,
    (\Pi_{U^\irp \perp} \otimes \mathbb{1})
    \bigg\{ \mathcal{F}_0^\irp
    \Big(
        \bigH^\irp
        + (\mathbb{1} \otimes \Pi_{U^\irp \perp}) \bigG^\irp
    \Big) \\ &+
    \Big(
        \sumB \mathcal{F}_0^{\irp\irpB} \mathcal{L}^\irpB
        + \sumBB \mathcal{F}_0^{\irp\irpB\,\irpBB} \mathcal{K}^{\irpB\,\irpBB}
        \Big)
    \Big(
        F_{I_0^\irp} \mathbf{\tilde{H}}_{I_0^\irp}
        + \matG_{I_0^\irp} \otimes (\Pi_{U^\irp \perp} \matG_{I_0^\irp})
    \Big)
    \bigg\}
  \end{split}
\end{equation}
\begin{equation}
  \begin{split}
    \Habsil_\irp^\irpP = &\,
    2 (\Pi_{U^\irp \perp} \otimes \Pi_{U^\irpP \perp})
    \Bigg\{ \mathcal{F}_0^{\irp\irpP}
    \bigg\{
    F_{I_0^\irpP} \matM^\irp \otimes \matG_{I_0^\irpP}
    + F_{I_0^\irp} \matG_{I_0^\irp} \otimes \matM^\irpP
     + \frac{1}{2} \sumijabmix \Dmix_{ij}^{ab, \irp\irpP}
      \hat{\matG}_{ia}^\irp \otimes \hat{\matG}_{jb}^\irpP\\
    & - \Big(
        \sum_{(i,a)\in\irp} \hat{\matG}_{ia}^\irp
        \sum_{(j,b)\in\irpP} F_{I_j^{b,\irpP}} \Dmix_{ij}^{ab, \irp\irpP}
    \Big) \otimes \matG_{I_0^\irpP}
    - \matG_{I_0^\irp} \otimes \Big(
        \sum_{(j,b)\in\irpP} \hat{\matG}_{jb}^\irpP
        \sum_{(i,a)\in\irp} F_{I_i^{a,\irp}} \Dmix_{ij}^{ab, \irp\irpP}
    \Big)
    \bigg\}\\
    & + 
    \Big(
        -C_0 \mathcal{F}_0^{\irp\irpP}
        + \sumBp \mathcal{F}_0^{\irp\irpP\irpB} \mathcal{L}^\irpB
        + \sumBBp \mathcal{F}_0^{\irp\irpP\irpB\,\irpBB} \mathcal{K}^{\irpB\,\irpBB}
    \Big)
    F_{I_0^\irp} F_{I_0^\irpP} \matG_{I_0^\irp} \otimes \matG_{I_0^\irpP}
    \Bigg\}
  \end{split}
\end{equation}
\end{widetext}
By inspection on these expressions one can see that the quantities $F_I$ and $\matG_I$ for single excitations,  $F_{I_i^{a,\irp}}$ and $\matG_{I_i^{a,\irp}}$,
are used often and their storage does not pose a problem.
Other quantities such as $\mathbf{\tilde{H}}_I$ and those associated to double excitations are too many for storage, but they are used only once.
Thus, an efficient implementation of these equations can be made that explores these facts.
Furthermore, the following relations among $F_I$, $\matG_I$ and $\mathbf{H}_I$ hold,
and can be used to calculate some of these quantities from others:
\begin{equation}
  F_{I_i^a} = (-1)^{i+n} \sum_{q=1}^n U_q^a \big( \matG_{I_0} \big)_q^i
\end{equation}
\begin{equation}
  F_{I_{ij}^{ab}} = (-1)^{i+n+(b>a)} \sum_{q=1}^n U_q^a \big( \matG_{I_j^b} \big)_q^i
\end{equation}
\begin{equation}
  \big( \matG_{I_i^a} \big)^p_q = \left\{
    \begin{array}{ll}
      (-1)^{i+n} \sum_{q'=1}^n (1-\delta_{qq'}) U_{q'}^a \big( \mathbf{H}_{I_0} \big)^{pi}_{qq'} & \quad p \ne a\\
      (-1)^{i+n} \big( \matG_{I_0} \big)^{i}_{q} & \quad p = a\\
    \end{array}
  \right.\,,
\end{equation}
where $(b>a) = 1$ if $b>a$, 0 otherwise.

\begin{acknowledgments}
  The authors thank the Dean's Office for Research of UFABC for providing the research facilities,
  and to the Coordena{\c c}{\~a}o de Aperfei{\c c}oamento de Pessoal de N{\'i}vel Superior - Brasil (CAPES) - Finance Code 001.
  Y.A.A. acknowledges the grants \#2017/21199-0 and \#2018/04617-6, S{\~a}o Paulo Research Foundation (FAPESP).
  Y.A.A. acknowledges useful discussions with Prof. Peter R. Taylor at conferences.
\end{acknowledgments}


%

\end{document}